\documentstyle[12pt,hyperref]{article}
\newcommand{\be}[1]{ \begin{equation}\label{#1} }
\newcommand{\ee}{\end{equation}}
\newcommand{\bea}[1]{\begin{eqnarray}\label{#1} }
\newcommand{\eea}{\end{eqnarray}}
\newcommand{\eq}[1]{(\ref{#1})}

\newcommand{\JJ}{{\cal J}}

\newcommand{\FF}{{\cal F}}
\newcommand{\NN}{{\cal N}}
\newcommand{\p}{\partial}
\newcommand{\wt}{\widetilde}
\def\ZZZ{{\hskip-3pt\hbox{ Z\kern-1.6mm Z}}}
\def\zzz{{\hskip-3pt\hbox{ z\kern-1mm z}}}

\def\ZZZ{{\hbox{ Z\kern-1.6mm Z}}}
\def\zzz{{\hbox{ z\kern-1mm z}}}

\newcommand{\eps}{\epsilon}

\newcommand{\vt}{\vartheta}

\newcommand{\ws}{{\wt\sigma}}
\newcommand{\wrh}{{\wt\rho}}
\newcommand{\wv}{{\wt v}}

\newcommand{\vu} {\vec u}
\newcommand{\vtau} {\vec \tau}
\newcommand{\htau} {\vec \eta}
\newcommand{\vj} {\vec J}
\newcommand{\vxi} {\vec \xi}
\newcommand{\vpsi} {\vec \psi}

\newcommand{\HH}{{\cal H}}

\newcommand{\CC}{{\cal C}}

\newcommand{\OO}{{\cal O}}

\newcommand{\wh}{\widehat}

\renewcommand{\wh}{\hat}

\newcommand{\vc}{\vec\chi}

\newcommand{\ben}{\begin{eqnarray}\displaystyle}
\newcommand{\een}{\end{eqnarray}}
\newcommand{\refb}[1]{(\ref{#1})}
\newcommand{\sectiono}[1]{\section{#1}\setcounter{equation}{0}}

\def\one{{\hbox{ 1\kern-.8mm l}}}
\def\zero{{\hbox{ 0\kern-1.5mm 0}}}

\begin{document}

{}~
{}~

 \hfill\vbox{\hbox{hep-th/0605210}
}\break

\vskip .6cm

{\baselineskip20pt
\begin{center}
{\Large \bf
CHL Dyons and Statistical Entropy Function
from D1-D5 System
}
\end{center} }

\vskip .6cm
\medskip

\vspace*{4.0ex}

\centerline{\large \rm
Justin R. David and Ashoke Sen}

\vspace*{4.0ex}

\centerline{\large \it Harish-Chandra Research Institute}

\centerline{\large \it  Chhatnag Road, Jhusi,
Allahabad 211019, INDIA}

\vspace*{1.0ex}

\centerline{\it E-mail: 
justin,sen@mri.ernet.in,
ashoke.sen@cern.ch}

\vspace*{5.0ex}

\vskip 1in

\centerline{\bf Abstract} \bigskip

We give a proof of 
the recently proposed formula for the
dyon spectrum in CHL string
theories by mapping it to a  configuration
of D1 and D5-branes and Kaluza-Klein monopole. We also 
give a prescription for computing the degeneracy as
a systematic  
expansion in inverse powers of charges. The computation can be
formulated as a problem of extremizing a duality invariant statistical
entropy function whose
value at the extremum gives the logarithm of the degeneracy. 
During this analysis we also determine the locations of the zeroes
and poles of the Siegel modular forms whose inverse give the dyon
partition function in the CHL models.

\vfill \eject

\baselineskip=18pt

\tableofcontents

\sectiono{Introduction} \label{sintro}

There exists a proposal for the exact degeneracy of dyons in
toroidally compactified heterotic string
theory\cite{9607026,0412287,0505094,0506249,0508174} 
and also in
toroidally compactified type II string theory\cite{0506151} in four
dimensions.
These
formul\ae\ are invariant under the S-duality transformations of the
theory, and also reproduce the entropy of a dyonic black hole in the
limit of large charges\cite{0412287}.
Ref.\cite{0505094} proposed a proof of this formula by first
relating this to a five dimensional D1-D5 system and then counting
the degeneracy of this system.  

In \cite{0510147,0602254,0603066}
this conjecture was generalized to a class of CHL
models\cite{CHL,CP,9507027,9507050,9508144,9508154}, obtained by
modding out heterotic string theory on $T^2\times T^4$ by a $\ZZZ_N$
transformation that involves $1/N$ unit of translation along one of
the circles of $T^2$ and a non-trivial action on the internal
conformal field theory (CFT) describing heterotic string
compactification on $T^4$. The values of $N$ considered in
\cite{0510147} were $N=$1,2,3,5,7, with $N=1$
representing toroidal compactification.  
The conjectured formul\ae\
for the dyon spectrum are invariant under the duality symmetries of the
theory, and also reproduce the black hole entropy for large charges
up to first non-leading order.

The goal of this paper is twofold. First we shall give a `proof' of the
conjectured formula by relating the dyons in the four dimensional 
CHL models to a D1-D5 system in five
dimensional CHL models. Although the basic idea behind
our analysis is similar
to that of \cite{0505094}, the details are different and even for the
$N=1$ case our proof does not reduce exactly to that given in 
\cite{0505094} (see \cite{0603066} for an attempt to generalize
the analysis of \cite{0505094} to CHL models).
The other goal will be to develop a systematic
procedure for extracting the asymptotic behaviour of the degeneracy
formula as a  series in inverse powers of charges and non-perturbative
corrections. During this analysis we find that the statistical entropy, 
defined as the logarithm of the degeneracy,
is obtained by extremizing a function. We call this function the 
statistical entropy function in analogy with the black hole entropy
function whose extremization gives the entropy of an extremal
black hole. The analogy in fact goes further since we show that up to
first non-leading order the statistical entropy function of the dyons 
matches the entropy function of extremal black holes carrying the same
charges.

Since the analysis involves a lot of technical details, we have organised
the paper such that only the basic ideas are presented in the 
text
and all technical details are relegated to the appendices. 
In \S\ref{sseq} we describe how the dyons in four dimensional
CHL models can be related
to a D1-D5 system in five dimensional CHL models. In  
\S\ref{sint} we  count the degeneracy of
the D1-D5 system under consideration and 
show that the result of this analysis
agreees with
the conjecture given in \cite{0510147,0602254}. Finally in 
\S\ref{sasymp} we develop a systematic procedure for extracting the
asymptotic behaviour of the degeneracy in the limit of large charges.

\sectiono{Dyons from D1-D5 System} \label{sseq}

In this section we shall follow the method of \cite{0505094} 
to relate the 
dyons in a four 
dimensional CHL model
to a rotating D1-D5 system in a five dimensional CHL 
model. 
The steps leading to this relation are as follows.

\begin{enumerate}
\item
Begin with type IIB string theory on 
$K3\times S^1$ with $Q_5$ D5-branes
wrapped on $K3\times S^1$, $Q_1$ D1-branes wrapped on $S^1$,
$-n$ units of momentum along $S^1$ and  angular 
momentum $J_1$ and $J_2$ 
in two independent planes with 
$J_1+J_2=J$\cite{9602065}.\footnote{If
we identify the tangent space group $SO(4)$ of the four transverse
spatial dimensions to $SU(2)_L\times SU(2)_R$, then $J$ can be
identified with twice the $U(1)_L$ generator of $SU(2)_L$.}
Since a K3-wrapped 
D5-brane carries $-1$ unit of D1-brane 
charge, this system has $Q_1-Q_5$ units of D1-brane charge along $S^1$.
For definiteness we shall choose the coordinate along $S^1$ such that
it has period $2\pi$.\footnote{Note that in our 
convention supersymmetry acts on the right-moving
sector of the D1-D5 world-volume theory, \i.e. on modes carrying
positive momentum along $S^1$. Thus a BPS state will involve only
excitations involving modes with negative or zero
momentum along $S^1$.}

\item
 Now place this system at the center of Taub-NUT space with
 coordinates of the Taub-NUT space transverse to $K3\times S^1$.
 The orientation of the Taub-NUT space is chosen 
such that the isometry 
direction of the Taub-NUT geometry that becomes 
the compact direction $\wt 
S^1$ in the 
asymptotic 
region coincides with the angular coordinate $\phi$ along which the black 
hole rotates\cite{0503217}. The new configuration now corresponds to a 
state in 
type IIB string theory on $K3\times S^1\times \wt S^1$ with $Q_5$ 
D5-branes wrapped on $K3\times S^1$, $(Q_1-Q_5)$ 
units of 
D1-brane charge along $S^1$,  $-n$ units of momentum along $S^1$,
a 
Kaluza-Klein monopole associated with the compact coordinate $\wt S^1$ and 
$J$ units of momentum along $\wt S^1$. 

\item
Make an S-duality transformation on this system to get type IIB string 
theory on
$K3\times S^1\times \wt S^1$ with
$Q_5$ NS5-branes on $K3\times S^1$, $(Q_1-Q_5)$ units of fundamental 
string winding charge along $S^1$, 
$-n$
units of momentum along $S^1$, $J$ units of momentum
along $\wt S^1$, and a Kaluza-Klein monopole associated with 
$\wt
S^1$ compactification.

\item
Now make an $R\to 1/R$ duality transformation along $\wt S^1$ to 
convert
the theory into IIA on 
$K3\times S^1\times \wh S^1$ with
$Q_5$ Kaluza-Klein monopoles associated with
$\wh S^1$ compactification, $(Q_1-Q_5)$ units of 
fundamental string winding charge
along 
$S^1$,
$-n$ units of momentum along $S^1$, $J$ units of 
fundamental string winding charge
along $\wh S^1$, and a single 
NS5-brane wrapped on
$K3\times S^1$. Here $\wh S^1$ 
denotes the dual circle of $\wt S^1$.

\item
Finally using the string-string 
duality\cite{9410167,9501030,9503124,9504027,9504047} 
relating heterotic on $T^4$ to 
type IIA string theory 
on $K3$, we can map this system to heterotic string theory on $T^4\times
S^1\times \wh S^1$ with
$Q_5$ Kaluza-Klein monopoles associated with 
$\wh S^1$ compactification, $(Q_1-Q_5)$ units of 
NS5-brane charge along
$T^4\times S^1$, $-n$ units of momentum along $S^1$,
$J$ units of NS5-brane charge along $K3\times \wh S^1$, and a single 
fundamental string
wrapped on
$S^1$.
If $Q_e$ and $Q_m$ denote the electric and magnetic charge vectors 
respectively, then this system has 
${1\over 2} \,
Q_m^2 = (Q_1-Q_5) Q_5 $, ${1\over 2}
Q_e^2 = n$, and $Q_e . Q_m = J$.
\end{enumerate}

Now we can mod out both sides of the duality by a
$\ZZZ_N$ transformation which, in the original theory, 
involves $2\pi/N$ translation along $S^1$
and some order $N$ transformation 
$\wt g$ acting on $K3$ that commutes with all the space-time
supersymmetry transformations. 
In the final heterotic string theory the $\ZZZ_N$ transformation acts
as $2\pi/N$ translation along $S^1$ together with some $\ZZZ_N$ action
on the coordinates of $T^4$ and the 16 left-moving internal 
bosonic coordinates.
In order to get a $\ZZZ_N$ invariant 
configuration so that we can carry out the $\ZZZ_N$ modding, we need 
to put periodic boundary conditions on all the 
branes which extend along $S^1$, and take $N$ identical copies of all the 
branes 
transverse to $S^1$ and place then at intervals of $2\pi/N$ along $S^1$. 
The latter 
set includes the five branes along $K3\times \wh S^1$;  we need to take 
$NJ$ five branes and place them at intervals of $2\pi / N$ along $S^1$. 
After orbifolding the 
direction along $S^1$ can be regarded as a circle of radius $1/N$, and we 
shall have $J$ five branes per unit period transverse to $S^1$. The 
natural unit of momentum along $S^1$ is now $N$, and momentum 
$-n$ along $S^1$ can be regarded as $-n/N$ units of momentum. The other 
charges have the same values as in the parent theory. This 
gives ${1\over 2}Q_e^2=n/N$. 
$Q_m^2$ and $Q_e\cdot Q_m$ remain unchanged from their original values. 
Thus we have
\be{echarges}
{1\over 2}Q_e^2=n/N, \qquad {1\over 2} Q_m^2 = (Q_1-Q_5)Q_5, \qquad
Q_e\cdot Q_m = J\, .
\ee
One point to note is that since we have
a single fundamental string along $S^1$, a $\ZZZ_N$ modding would correspond 
to 
requiring that under a shift of $2\pi/N$ along $S^1$ the non-compact 
coordinates of the string transverse to $S^1$ come back to their originl 
values and the coordinates of the string along $K3$ come back to the image 
of the 
original location under the $\ZZZ_N$ generator $\wt g$ acting on K3. This 
represents a twisted sector state in the orbifold theory.
This is consistent with the assertion of \cite{0510147} that the proposed 
answer for the degeneracy of dyonic states is
valid only for states which carry electric charges compatible with twisted 
sector states.

This analysis shows that the problem of computing degeneracy of dyons
in the
four dimensional CHL model can be reduced to the problem of 
computing the degeneracy of a rotating D1-D5 system in Taub-NUT
space, carrying momentum along the direction common to the D1-brane
and the D5-brane.
We shall now turn to the problem of computing degeneracy of this
system.

\sectiono{Counting of States of the Rotating D1-D5 System in
Taub-NUT Space} 
\label{sint}

In this section we shall compute the degeneracy of the 
D1-D5 system described in the previous section.
For simplicity we shall
restrict our analysis to the $Q_5=1$ case,   --
the generalization of this 
analysis to the $Q_5\ne 1$ case has been discussed in 
appendix \ref{ssym}. 
In this case 
an intuitive picture of the states of the D1-D5 system may be given as 
follows\cite{9512078}.
First of all dynamics of the 1-5 strings produces an effective binding
between the D1 and D5 which forces the 
D1-branes to move in the plane
of the D5-brane. Thus the only zero modes of the D1-brane
associated with the
transverse directions are those along K3. A generic state of the system
will contain a certain number of isolated D1-branes in the
plane of the D5-brane, with the $i$-th
D1-brane wound $w_i$ times along $S^1$ and
carrying momenta $-l_i$ along $S^1$
and  $j_i$ along $\wt S^1$, with $w_i,l_i,j_i\in\ZZZ$. 
The BPS condition will also require $w_i>0$, $l_i\ge 0$. The overall
motion of the D1-D5-system in Taub-NUT space can carry some additional
momentum $-l_0$ along $S^1$ and $j_0$ along $\wt S^1$.
Furthermore the low energy
dynamics of closed string modes localized near the core of the
Taub-NUT space can also carry some momentum $-l_0'$ along
$S^1$.\footnote{Localized low 
energy excitations  around the
Taub-NUT space  do not carry any momentum along $\wt S^1$.
They can carry fundamental string winding charge 
along $\wt S^1$\cite{9705212},
but for the configuration which is of interest to us, these charges
vanish.}
Thus we
have the constraint
\be{esi1}
\sum_i w_i = Q_1, \qquad l_0'+l_0+
\sum_i l_i = n, \qquad j_0+ \sum_i j_i=J\, .
\ee
Each of the systems described above
must satisfy the  boundary
condition that under a translation along $S^1$ by $2\pi/N$
its oscillation modes along $K3$ get twisted by $\wt g$.
 
Our goal is to compute the degeneracy of the
D-brane system described above. In fact what we shall really
compute is not the degeneracy, but an appropriate index defined
as follows. First of all the D1-D5-Kaluza Klein 
monopole system in type IIA on $K3\times S^1\times \wt S^1$ 
breaks
twelve of the sixteen supercharges of the bulk theory. Thus there will
be twelve fermionic zero modes associated with the broken 
supersymmetry
generators. Quantization of these fermion zero modes produce a 
$2^6=64$-fold degeneracy of states. These states form a single
irreducible multiplet of the supersymmetry algebra known as the
intermediate multiplet containing equal number of fermionic and
bosonic states. This is then tensored with the states obtained
by quantizing the rest of the degrees of freedom of the theory to get
the full spectrum. If the state that is tensored with the basic supermultiplet
is bosonic we shall call this a bosonic supermultiplet and
if this state is fermionic we
shall call this a fermionic supermultiplet. In other words we shall call a
supermultiplet bosonic (fermionic) if the highest spin state in the
supermultiplet is bosonic (fermionic). We shall 
denote by $h(Q_1,n,J)$ the number
of bosonic supermultiplets minus the number of fermionic supermultiplets 
of the
system described above carrying quantum numbers $(Q_1,n,J)$.
By an abuse of notation we shall continue to refer to this number as
the degeneracy of states.

We shall first analyze
the degeneracy $d_{KK}(l_0')$ associated with the low energy
dynamics of  the taub-NUT space, carrying
momentum $-l_0'$ along $S^1$. 
Under the duality transformation described in section
\ref{sseq} the Taub-NUT space gets mapped to a fundamental
string wound once along $S^1$ and carrying momentum
$-l_0'$ along $S^1$ in heterotic string theory on 
$(T^4\times S^1)/\ZZZ_N
\times
\wh S^1$. Since this is a twisted sector state, the problem of
computing $d_{KK}(l_0')$
reduces to the problem of computing the 
degeneracy of twisted sectors states in this string theory. This has
been done in appendix \ref{stwisted} and the result is 
\be{etwo}
\sum_{l_0'} d_{KK}(l_0') e^{2\pi i l_0' \wh\rho} 
= 16\, e^{-2\pi i \wh\rho} \prod_{n=1}^\infty \left\{
(1 - e^{2\pi i n\wh\rho})^{-{24\over N+1}}
(1 - e^{2\pi i nN\wh\rho})^{-{24\over N+1}}\right\}
\ee
The Taub-NUT space breaks eight of the sixteen supersymmetry
generators in type IIB string theory on K3. This gives rise
to eight fermonic zero modes.
The factor of 16 in \refb{etwo} arises from the quantization of these
zero modes.

 Next we turn to
the degeneracy $d_{CM}(l_0,j_0)$
 associated with the
overall motion of the D1-D5-system in Taub-NUT space carrying
momentum $-l_0$ along $S^1$ and $j_0$ along $\wt S^1$.
The low energy theory describing
the overall dynamics of the D1-D5 system in Taub-NUT space
is that of a supersymmetric $\sigma$-model with
Taub-NUT target space, the coordinate along
$S^1$  being 
identified as the world-sheet $\sigma$ coordinate of this
supersymmetric field theory. 
Since $S^1/\ZZZ_N$ has period $2\pi/N$,
the natural unit of momentum along
$\sigma$ is $N$. As a result a BPS state
of the D-brane system carrying momentum $-l_0$ correponds
to $L_0=l_0/N$, $\bar L_0=0$.\footnote{Throughout this paper 
$L_0$ and $\bar L_0$ of a state in a CFT will denote the left and
right-moving world-sheet momenta with a normalization in which
the world-sheet coordinate $\sigma$ has period $2\pi$. In RR sector
we implicitly subtract the factors of $c/24$ ($\bar c/24$) from
$L_0$ ($\bar L_0$) which
arise in going from the sphere to the cylinder coordinates.}
On the other hand
the momentum $j_0$ along
$\wt S^1$ is the U(1) charge associated with the angular direction
of the Taub-NUT space
which becomes the compact circle asymptotically.
The
degeneracy of such states has been computed in appendix \ref{sb},
and the result is
\bea{eone}
&& \sum_{l_0, j_0} d_{CM}(l_0, j_0) e^{2\pi i l_0\wh\rho + 2\pi i
j_0\wh v} = 4 \, e^{-2\pi i \wh v} \,  (1 - e^{-2\pi i \wh v})^{-2}\,
\nonumber \\
&& \qquad \qquad \prod_{n=1}^\infty \left\{
(1 - e^{2\pi i n N\wh\rho})^4 \, ( 1 - e^{2\pi i n N\wh\rho + 2\pi i
\wh v})^{-2} \,  ( 1 - e^{2\pi i n N\wh\rho - 2\pi i
\wh v})^{-2}\right\}\, . \nonumber \\
\eea
The D1-D5 system in K3$\times$Taub-NUT space breaks four of the
eight supersymmetry generators of type IIB string theory on $K3\times 
\hbox{Taub-NUT}$. Quantization of the associated
zero modes gives rise to the factor of 4 in \refb{eone}.
The factor of 16 in \refb{etwo} and 4 in \refb{eone}  together 
provides the 64-fold degeneracy of a 1/4 BPS supermultiplet.
As pointed out earlier, these 64 states contain equal numbers of
bosons and fermions. After factoring this out we count the rest of
the states with weight +1 for bosons and weight $-1$ for fermions.

Let us now turn to the computation of the degeneracy $n(w,l,j)$ 
of a single D1-brane
moving inside the D5-brane,
carrying 
winding $w$,  
momentum $-l$  and 
angular momentum $j$. 
We denote by $\sigma$ the coordinate along the length of the
D1-brane, $\sigma$ being normalized so that it coincides with the
target space coordinate in which the original $S^1$ had period
$2\pi$. Since the
D1-brane carries winding charge $w$, $\sigma$ changes by $2\pi w/N$
as we traverse the whole length of the string, regarded as a configuration
in the orbifold.
Under $\sigma\to \sigma+ 2\pi w/N$, the physical coordinate
of the D1-brane shifts by $2\pi r / N$  along $S^1$ where
\be{esi2}
r=w \quad \hbox{mod $N$}\, .
\ee
$\ZZZ_N$ invariance then
requires that under $\sigma\to \sigma+ 2\pi w/N$ the location
of the D1-brane along K3 gets transformed by $\wt g^r=\wt g^w$.

We expect the low energy dynamics of this D-brane system to be
described by a superconformal field theory  (SCFT)
with target space $K3$ subject to the above boundary condition.
Since the D1-brane is subject to twisted boundary condition
with period $2\pi w/N$, the natural unit of momentum on the
brane is $N/w$. Thus a total momentum $-l$ corresponds to $-lw/N$
unit of momentum and can be identified with the $\bar L_0-L_0$
eigenvalue of the state. Since the BPS condition on the
D1-brane corresponds to $\bar
L_0=0$ we have $L_0=lw/N$.
Thus we are looking for a state in the SCFT with
\be{esi3}
L_0=lw/N, \qquad \bar L_0 = 0\, .
\ee
The bosonic and fermionic excitations on the brane satisfy identical
boundary conditions in order to preserve space-time supersymmetry.
Hence the state belongs to the Ramond-Ramond (RR) sector.
Furthermore the boundary condition described below
eq.\refb{esi2} tells us that
the state is twisted by $\wt g^r$. 
Finally, since the total momentum along
$S^1$ is $-l$, under translation by $2\pi /N$
along $S^1$ this state picks up a phase $e^{-2\pi i l/N}$. Thus the
projection operator onto $\ZZZ_N$ invariant states is
given by
\be{esi6}
{1\over N}\, \sum_{s=0}^{N-1} e^{-2\pi i sl/N} \wt g^s \, .
\ee
Putting these results together we see that the total number of
$\ZZZ_N$ invariant bosonic minus fermionic
states of the single D1-brane carrying quantum
numbers $w,l,j$ is given by
\be{esi7}
n(w,l,j) \equiv {1\over N}
\, \sum_{s=0}^{N-1} e^{-2\pi i sl/N}
Tr_{RR;\wt g^r} \left(\wt g^s(-1)^{F+\bar F} \delta_{NL_0,lw}
\delta_{{\cal J}, j}\right) \, .
\ee
where $Tr_{RR;\wt g^r}$ denotes trace over RR sector states
twisted by $\wt g^r$, $F,\bar F$ are world-sheet fermion numbers
(which coincide with the space-time fermion numbers) associated
with the left and right-moving excitations on the D1-brane, and
${\cal J}$ is the $U(1)$ generator associated with an appropriate
R-symmetry current of this conformal field 
theory\footnote{In this
 case the full R-symmetry group is 
 $SO(4)\simeq SU(2)_L\times SU(2)_R$
 and $\JJ$ corresponds to twice the  $U(1)_L$ generator of 
 the $SU(2)_L$ group.}
which can be
identified with the 
 angular momentum operator\cite{9602065}. Insertion of
 $(-1)^{\bar F}$ in the trace automatically projects 
 onto $\bar L_0=0$
 states.

Let us define\cite{0602254}
\be{esi4a}
F^{(r,s)}(\tau,z) \equiv {1\over N} Tr_{RR;\wt g^r} \left(\wt g^s
(-1)^{F+\bar F}
e^{2\pi i \tau L_0} e^{2\pi i {\cal J} z}\right) 
\ee
Then we have\cite{0602254}
\bea{fifth}
F^{(0,0)}(\tau, z) &=& {8\over N} A(\tau, z)\, ,
\nonumber \\
F^{(0,s)}(\tau, z) &=& {8\over N(N+1)} \, A(\tau, z) -{2\over N+1}
\, B(\tau, z) \, E_N(\tau) \qquad \hbox{for $1\le s\le (N-1)$}
\, , \nonumber \\
F^{(r,rk)}(\tau, z) &=& {8\over N(N+1)} \, A(\tau, z)
+ {2\over N(N+1)} \, E_N\left({\tau+k\over N}\right)\, B(\tau, z)\, ,
\nonumber \\
&& \qquad \qquad \qquad \qquad 
\qquad \hbox{for $1\le r \le (N-1)$, $0\le k\le (N-1)$}
\, ,\nonumber \\
\eea
where 
\be{efirst}
A(\tau, z) =  \left[ {\vartheta_2(\tau,z)^2
\over \vartheta_2(\tau,0)^2} +
{\vartheta_3(\tau,z)^2\over \vartheta_3(\tau,0)^2}
+ {\vartheta_4(\tau,z)^2\over \vartheta_4(\tau,0)^2}\right]\, ,
\ee
\be{second}
B(\tau, z) = \eta(\tau)^{-6} \vartheta_1(\tau, z)^2\, ,
\ee
and
\be{third}
E_N(\tau) = {12 i\over \pi(N-1)} \, \p_\tau \left[ \ln\eta(\tau)
-\ln\eta(N\tau)\right]= 1 + {24\over N-1} \, \sum_{n_1,n_2\ge 1\atop
n_1 \ne 0 \,  mod \, N} n_1 e^{2\pi i n_1 n_2 \tau}\, .
\ee
$F^{(r,s)}(\tau, z)$ has power series expansion of the 
form\cite{0602254}
\be{esi4b}
F^{(r,s)}(\tau,z)= \sum_{n\in \zzz
/N,j\in\zzz} c^{(r,s)}(4n - j^2) e^{2\pi in\tau} 
e^{2\pi i  j z}\, ,
\ee
for appropriate coefficients $c^{(r,s)}(4n - j^2)$.
{}From  
\refb{esi4a}, \refb{esi4b} it follows that
 \be{esi5}
{1\over N}\,
tr_{RR,\wt g^r} \left(\wt g^s(-1)^{F+\bar F}\delta_{NL_0,lw}
\delta_{{\cal J}, j}
\right) = c^{(r,s)}(4lw/N - j^2)\, .
\ee
Hence \refb{esi7} gives
\be{esi7.1}
n(w,l,j)  
= 
\sum_{s=0}^{N-1} e^{-2\pi i sl/N} c^{(r,s)}(4lw/N - j^2)\, ,
\qquad \hbox{$r=w$ mod $N$}\, .
\ee

Using the result for single D1-brane spectrum, we can now evaluate
the degeneracy of multiple D1-branes moving inside the D5-brane. Let
$d_{D1}(W,L,J')$ denote the degeneracy of this system, carrying total
D1-brane charge $W=\sum_i w_i$, total momentum $-L=-\sum_i l_i$
along $S^1$ and total angular momentum $J'=\sum_i j_i$. Then a
straightforward combinatoric analysis shows that
\be{emulti}
\sum_{W,L,J'} d_{D1}(W,L,J') e^{2\pi i ( 
\wh \sigma W /N +\wh \rho L + \wh v J')}
= \prod_{w,l,j\in \zzz\atop w>0, l\ge 0}  
\left( 1 - e^{2\pi i (\wh \sigma w / N + \wh \rho l + 
\wh v j)}\right)^{-n(w,l,j)}\, .
\ee

Let us now turn to the full system containing the D5-brane
and multiple D1-branes in the Kaluza-Klein monopole
background.
The combinatoric problem to be solved is the following. We have
total quantum numbers $(Q_1, n, J)$, to be distributed among
multiple D1-branes moving inside a D5-brane, the overall
D1-D5 system and the Taub-NUT space
according to the relation \refb{esi1}. The degeneracy associated with the
dynamics of (multiple) D1-branes inside the D5-brane, 
carrying quantum numbers $(W=\sum w_i,L=\sum l_i,J'
=\sum j_i)$
is given by $d_{D1}(W,L,J')$, the
degeneracy associated with the overall dynamics of the D1-D5
system carrying quantum numbers $l_0$,
$j_0$  is given by $d_{CM}(l_0, j_0)$ and the degeneracy associated
with the dynamics of the
Taub-NUT space carrying quantum number $l_0'$
is given by $d_{CM}(l_0')$.  If $h(Q_1, n,J)$ denotes
the total number of bosonic minus fermionic 
supermultiplets (in the sense described earlier)
of the mutiple D1-brane system carrying
quantum numbers $(Q_1,n,J)$ and if we define
\be{esi8}
f(\wh \rho,\wh \sigma,\wh v) = \sum_{Q_1, n, J} h(Q_1,n, J)
e^{2\pi i ( \wh \rho n + \wh \sigma (Q_1-1) /N +\wh v J)} \, ,
\ee
then we get
\bea{esi9}
f(\wh \rho,\wh \sigma,\wh v ) &=& {1\over 64} 
\, e^{-2\pi i \wh \sigma /N}\, \sum_{W,L,J'} d_{D1}(W,L,J') 
e^{2\pi i ( 
\wh \sigma W /N +\wh \rho L + \wh v J')}
\nonumber \\
&& \, \left(\sum_{l_0, j_0} 
d_{CM}(l_0, j_0) e^{2\pi i l_0\wh\rho + 2\pi i
j_0\wh v}\right) \, 
\left(\sum_{l_0'} d_{KK}(l_0') e^{2\pi i l_0' \wh\rho} \right)\, .
\nonumber \\
\eea
The factor of $1/64$ in this equation arises due to the fact that a single
1/4 BPS supermultiplet is 64-fold degenerate; thus in order to count
the number of supermultiplets we need to divide the total number of
states by 64.
Using \refb{etwo}, \refb{eone}, \refb{esi7.1}, \refb{emulti}  
and the 
relations\cite{0602254}
\bea{ecrsv}
&& c^{(0,0)}(0) = {20\over N}\, , \qquad 
c^{(0,0)}(-1) = {2\over N}\, , \nonumber \\
&& c^{(0,s)}(0) = {1\over 
N}\left( 20 - {24 
N\over N+1}\right)\, , 
\qquad c^{(0,s)}(-1) = {2\over N}\, , \quad \hbox{for 
$s=1,2, \ldots (N-1)$}\, , \nonumber \\
\eea
one can reduce \refb{esi9} to 
\be{esi9ex}
f(\wh \rho,\wh \sigma,\wh v )= e^{-2\pi i (\wh\rho + \wh \sigma/N
+ \wh v)}
\prod_{r=0}^{N-1}
\prod_{k'\in \zzz+{r\over N},l,j\in \zzz\atop k',l\ge 0, j<0 \, {\rm for}
\, k'=l=0}
\left( 1 - e^{2\pi i (\wh \sigma k'   + \wh \rho l + \wh v j)}\right)^{
-\sum_{s=0}^{N-1} e^{-2\pi i sl/N } c^{(r,s)}(4lk' - j^2)}\, .
 \ee
The $k'=0$ term in the last expression comes from the terms involving
$d_{CM}(l_0, j_0)$ and $d_{KK}(l_0')$.  
 Comparing the right hand side of this equation with the expression
 for $\wt\Phi_k$ given in \refb{es05a}
we can rewrite \refb{esi9} as
\be{esi9a}
f(\wh \rho,\wh \sigma,\wh v )
= -{(i\sqrt N)^{-k-2}
\over \wt\Phi_k(\wh \rho,\wh \sigma, \wh v)}\, .
\ee
$\wt\Phi_k$ is a weight $k$ modular form of an appropriate
subgroup of
the Siegel modular group\cite{0602254}.

{}From \refb{esi8}, \refb{esi9a} it follows that for $n>0$,
$Q_1>1$, $J>0$,
\bea{esi11pre}
h( Q_1, n,J) &=&  {K}
\int_{\wh C} d\wh \rho d\wh\sigma d\wh v
{1\over \wt\Phi_k(\wh \rho, \wh \sigma, \wh v)}
\exp\left[ -2\pi i(\wh \rho n+ \wh \sigma (Q_1-1)/N +
\wh v J)\right]\, , \nonumber \\
\een
where 
\be {edefk}
K = -{1\over N} (i\sqrt N)^{-k-2}\, ,
\ee
and the integration surface $\wh C$ is the
three real dimensional surface:
\bea{ep2aa}
Im\, \wh \rho=M_2/N, \quad Im \, \wh\sigma = M_1N, \quad
Im \, \wh v = M_3, \nonumber \\
 0\le Re\, \wh\rho\le 1, \quad
0\le Re\, \wh\sigma\le N, \quad 0\le Re\,  \wh v\le 1\, ,
\een
$M_1$, $M_2$ and $M_3$ being three large but fixed positive
numbers.

Now recall that from the  map described in section
\ref{sseq}, we have for this system
\be{esi10}
Q_e^2 = 2n/N, \qquad Q_m^2 = 2 (Q_1-Q_5) Q_5=
2 (Q_1-1), \qquad
J = Q_e\cdot Q_m\, .
\ee
Thus the degeneracy $d(Q_e, Q_m)$ of the four 
dimensional black hole
is related to the quantity $h( Q_1 , n,J)$ given via 
eqs.\refb{esi8}, \refb{esi9a} as
\bea{esi11}
d(Q_e, Q_m) &=& h\left({1\over 2} {Q_m^2}+1,
{1\over 2} Q_e^2 N, 
Q_e\cdot Q_m\right)\nonumber \\
&=& K
\int_{\wt C} d\wh \rho d\wh\sigma d\wh v
{1\over \wt\Phi_k(\wh \rho, \wh \sigma, \wh v)}
\exp\left[ -i\pi (\wh \rho Q_e^2 N+ \wh \sigma Q_m^2/N +
2\wh v Q_e\cdot Q_m)\right]\, . \nonumber \\
\een

It has been shown in appendix \ref{sproof} that 
\be{epr1}
\wt\Phi_k(\wh \rho, \wh \sigma, \wh v) 
= \wt\Phi_k(\wh \sigma/N, \wh \rho N, \wh v)\, .
\ee
Thus defining new variables:
\be{epr2}
\wt\rho = \wh \sigma/N, \quad \wt\sigma=\wh \rho N, \quad \wv=
\wh v, 
\ee
we can write 
\bea{ep1}
d(Q_e, Q_m) &=&  K
\int_C d\wt \rho d\wt\sigma d\wt v
{1\over \wt\Phi_k(\wt \rho, \wt \sigma, \wt v)}
\exp\left[ -i\pi (\wt \rho Q_m^2+ \wt \sigma Q_e^2  +
2\wt v Q_e\cdot Q_m)\right]\, , \nonumber \\
\een
with the integration surface $C$ being the
three real dimensional surface:
\bea{ep2}
Im\, \wt \rho=M_1, \quad Im \, \wt\sigma = M_2, \quad
Im \, \wt v = M_3, \nonumber \\
 0\le Re\, \wt\rho\le 1, \quad
0\le Re\, \wt\sigma\le N, \quad 0\le Re\,  \wt v\le 1\, .
\een
This agrees with the proposal put forward in 
\cite{0510147,0602254,0603066}.

Although this formula has been derived for $Q_5=1$, a more systematic 
analysis based on mapping the low energy dynamics of the D1-D5 system 
to a conformal field theory with the symmetric product of 
$(Q_1-Q_5) Q_5+1$ copies of $K3$ as the 
target space\cite{9512078}
can be used to show that the final formula holds 
for general $Q_5$. This has been done in appendix \ref{ssym}.

\sectiono{Asymptotic Behaviour of the Degeneracy of Dyons}
\label{sasymp}

In this section we shall develop a systematic method for
computing the behaviour of  $d(Q_e, Q_m)$ given in \refb{ep1} 
for large charges.
As in \cite{9607026,0412287,0510147,0601108}
we shall try to estimate the behaviour of this integral
for large charges by deforming the integration contour and picking up
residues from the poles. 
This removes one of the three integrals. The remaining two integrals
will then be performed using a saddle point approximation. At the end
of this process we need to compare the contributions from different
poles and identify the one that gives the dominant contribution in the
large charge limit.

The poles of the integrand in \refb{ep1}
arise from the zeroes of $\wt\Phi_k$. The locations of the
zeroes and poles of $\wt\Phi_k$ have  been determined in
appendix \ref{szero}. 
According to eqs.\refb{ev2}, \refb{ev4}, 
$\wt\Phi_k$ has second order zeroes at
\bea{ep4}
&&  n_2 ( \ws \wrh  -\wv ^2) + b\wv  
+ n_1 \ws  -\wrh m_1 + m_2
 = 0, \nonumber \\
&& \hbox{for} \quad  \hbox{$m_1\in N\ZZZ$, 
$n_1\in\ZZZ$,
$b\in 2\ZZZ+1$, $m_2, n_2\in \ZZZ$}, \quad
m_1 n_1 + m_2 n_2 +\frac{b^2}{4} = {1\over 4}\, .
\nonumber \\
\eea
Let us define
\be{ep3}
A = n_2, \quad B = (n_1, -m_1, {1\over 2} b), 
\quad y = (\wt\rho, \wt\sigma, 
-\wt v), \qquad
C = m_2\, , \quad q = (Q_e^2, Q_m^2,  Q_e\cdot Q_m)\, ,
\ee
and denote by $\cdot$ the $SO(2,1)$ invariant inner product
\be{ep5}
(x^1, x^2, x^3)\cdot (y^1, y^2, y^3) = x^1 y^2 + x^2 y^1 - 2 x^3 y^3\, .
\ee
Then we have
\be{eex}
y^2\equiv y\cdot y = 2(\wt\rho\wt\sigma - \wt v^2), \qquad
B\cdot y = b\wt v + n_1\wt\sigma - m_1\wt\rho\, ,
\ee
and the first equation of \refb{ep4} may be rewritten as
\be{efourtwo}
{1\over 2} A y^2 + B\cdot y + C 
=0\, .
\ee
 Picking up residue at the pole forces us to evaluate the exponent
 in \refb{ep1}
 \be{ep6}
 -i\pi\left(\wt \rho Q_m^2 + \wt \sigma Q_e^2 +
2\wt v Q_e\cdot Q_m\right) = -i\, \pi \, q \cdot y\, ,
\ee
at \refb{efourtwo}. To leading approximation
the location of the saddle point is now determined by
 extremizing \refb{ep6} with respect to $y$
subject to the condition \refb{efourtwo}. This gives
\be{ep7}
q + \lambda (A y + B) = 0\, ,
\ee
where $\lambda$ is a lagrange multiplier. 
\refb{efourtwo} and \refb{ep7}
now give:
\be{ep8}
\lambda = \pm
\sqrt{q^2 \over B^2 - 2 A C}, \qquad y = -{1\over A} \left(
{q\over \lambda} + B\right)\, .
\ee 
Since
\be{eq1}
B^2 - 2 A C = - 2 (m_1 n_1 + m_2 n_2 + {b^2\over 4}) 
= -{1\over 2}
\ee
due to the last equation in \refb{ep4}, we get
\be{eq3}
 \lambda = \pm\sqrt{-2q^2}\, ,
 \ee
Thus at the saddle point the exponential $e^{-i\pi q\cdot y}$
takes the form:
\be{eq2}
E \equiv  e^{-i\pi\, q\cdot y} =
e^{i\pi (q^2 / A \lambda + q\cdot B/A)}= e^{\left(
\pm\pi \sqrt{q^2 / 2}
+ i\pi q\cdot B\right)/A}\, .
\ee
 Now since $q\cdot B$ and A are integers, 
 the second term only gives a phase.
Hence
 \be{eq4}
 |E| = e^{{\pi\over A} \sqrt{q^2 / 2}  } = e^{\pi \sqrt{Q_e^2 Q_m^2 
 - (Q_e\cdot Q_m)^2} / n_2}\, .
 \ee
 Note that we have chosen the sign of the square root so that
 the sign in the exponent is positive since this gives the dominant 
 contribution.
 
 Eq.\refb{eq4} shows that the leading contribution to the integral
 comes from the saddle point corresponding to $n_2=1$. In this case
 a $\wrh \to \wrh +1$ transformation in \refb{ep4}
 induces $n_1\to n_1+1$, $m_2\to 
 m_2-m_1$, and since $n_1\in \ZZZ$,
 we can use this symmetry to bring the saddle point
 to $n_1=0$. On the other hand a $\ws \to \ws +N$
 transformation in eq.\refb{ep4}
 induces $m_1\to m_1-N$, $m_2\to m_2+n_1N$. Since $m_1\in
 N\ZZZ$, we can use this transformation to bring $m_1$ to 0.
 Finally the $\wv \to \wv +1$ transformation in \refb{ep4} induces
 $b\to b-2$, $m_2\to m_2 +b- 1$. Since $b\in 2\ZZZ+1$, we can use this
 transformation to set $b=1$. $m_2$ is now determined to be zero
 from the last equation in \refb{ep4}. Thus we have
 \be{eq5}
 m_1=m_2=n_1=0, \quad n_2 = 1, \quad b=1\, .
 \ee
 The corresponding zero of $\wt\Phi_k$ is at
 \be{eq6}
 \ws \wrh  - \wv ^2 +\wv  = 0\, .
 \ee
 
 The exponent of eq.\refb{eq4} for $n_2=1$ 
 gives the leading term in the expression for the
 statistical entropy, but in order to find a systematic expansion of
 the entropy in inverse power of charges, we need to carefully
 evaluate the complete contribution from the pole at \refb{eq6}.
 For this it will be useful to
 define new variable $\rho$, $\sigma$, $v$ through the
relations:
\be{e3.8}
   \rho = {\wt \rho \wt\sigma - \wt v^2\over \wt\sigma}, 
   \qquad \sigma = {\wt\rho \wt \sigma - (\wt v - 1)^2\over \wt 
   \sigma}
   \, , \qquad
   v = {\wt\rho \wt\sigma - \wt v^2 + \wt v\over \wt\sigma}\, ,
\ee
or equivalently,
\be{e3.8a}
   \wt\rho = {v^2-\rho\sigma \over 2v-\rho-\sigma}, \qquad
   \wt\sigma={1\over 2v-\rho-\sigma}, \qquad \wt v =
   {v-\rho \over 2v-\rho-\sigma}\, .
\ee
In these variables \refb{eq6} gets mapped to $v=0$, and we have,
near $v=0$\cite{0510147}, 
\bea{eq8}
\wt \Phi_k(\wt\rho,\wt\sigma,\wt v) &=& 4\pi^2\, (2v -\rho-\sigma)^{k}
\, v^2\, f^{(k)}(\rho) f^{(k)}(\sigma) + \OO(v^4)\, ,
\nonumber \\
f^{(k)}(\tau) &\equiv& \eta(\tau)^{k+2} \eta(N\tau)^{k+2}\, .
\eea
Also it follows from \refb{e3.8}, \refb{e3.8a} that
\be{eq9}
d\wrh   d\ws   d\wv = (2v - \rho - \sigma)^{-3}\, 
d\rho   d\sigma  dv\, .
\ee
Thus the contribution to \refb{ep1} from the pole at $\wt\rho 
\wt\sigma - \wt v^2 + \wt v=0$ is given by
\bea{er1}
d(Q_e, Q_m)&\simeq& 
{K\over 4\pi^2}\, \int_{C'}  d\rho   d\sigma  
dv \, v^{-2} \, (2v -\rho-
\sigma)^{-k-3} \, \left(f^{(k)}(\rho) f^{(k)}(\sigma)
\right)^{-1} \nonumber \\
&& \exp\left[ -i\pi \left\{ {v^2 -\rho\sigma\over 2v 
-\rho-\sigma}
Q_m^2 +{1\over 2v -\rho-\sigma} Q_e^2 +{2(v-\rho)\over 2v -\rho-
\sigma} Q_e\cdot Q_m\right\}\right]\, , \nonumber \\
\een
where the integration contour $C'$ now encloses the divisor $v=0$.
The correction to this formula involves contribution from other
poles for which $n_2\ne 1$, and are
suppressed by powers of $e^{-Q^2}$.
Evaluating the $v$ integral in \refb{er1}
by Cauchy's formula, we get
\bea{er2}
d(Q_e, Q_m) &\simeq& i{K\over 2\pi} \, 
(-1)^{k+Q_e\cdot Q_m} \, \int 
{d\rho   d\sigma\over (\rho+\sigma)^2} \nonumber \\ && 
\left[ -2 (k+3) + 2 \pi i \, \left\{ {\rho\sigma\over \rho+\sigma} Q_m^2
-{1\over \rho+\sigma} Q_e^2 + {\rho-\sigma\over \rho+\sigma}
Q_e\cdot Q_m\right\} \right]\nonumber \\
&&\exp\Bigg[ -i\pi \left\{ {\rho\sigma\over \rho+\sigma} Q_m^2
-{1\over \rho+\sigma} Q_e^2 + {\rho-\sigma\over \rho+\sigma}
Q_e\cdot Q_m\right\}  \nonumber \\
&& -\ln f^{(k)}(\rho)-\ln f^{(k)}(\sigma)
- (k+2) \ln (\rho+\sigma)\Bigg]\, . \nonumber \\
\een

Let us now introduce new complex variables $a$ and $S$ through the
relations:
\be{er3}
\rho = a+iS, \qquad \sigma = -a+iS\, .
\ee
Then \refb{er2} may be rewritten as
\bea{er4}
d(Q_e, Q_m) &\simeq& -{K\over 4\pi} \, 
(-1)^{k+Q_e\cdot Q_m} \, \int {dS   da \over S^2} \nonumber \\
&&
\left[ 2(k+3) + {\pi\over S} \left\{ (a^2 + S^2)  Q_m^2 +  Q_e^2
- 2a Q_e\cdot Q_m\right\} \right]\nonumber \\
&& \exp\Bigg[ {\pi\over 2 S} \left\{ (a^2 + S^2)  Q_m^2 +  Q_e^2
- 2a Q_e\cdot Q_m\right\} \nonumber \\ &&
-\ln f^{(k)}(a + iS) -\ln f^{(k)}(-a + iS)
- (k+2) \ln (2 i S)\Bigg]\, .
\een

So far we have not specified the contour for $a$ and $S$ integration,
except that it must pass through the saddle point. In the leading
approximation the saddle point, obtained by extremizing
\be{er5}
{\pi\over 2 S} \left\{ (a^2 + S^2)  Q_m^2 +  Q_e^2
- 2a Q_e\cdot Q_m\right\}
\ee
occurs for real values of $a$ and $S$. Thus we can take the 
contour of integration to be along the real $a$ and $S$ axis.
If we now define
\be{er6}
\tau = -a + i S \equiv \tau_1 + i \tau_2\, ,
\ee
then \refb{er4} may be  reexpressed as
\bea{er7}
d(Q_e, Q_m) &\simeq& K_0 (-1)^{Q_e\cdot Q_m} \, 
\int {d^2 \tau\over \tau_2^2}\, 
\left[ 2(k+3) + {\pi\over \tau_2} |Q_e +\tau Q_m|^2\right] \nonumber \\
&& \exp\Bigg[ {\pi\over 2 \tau_2} \, |Q_e +\tau Q_m|^2
-\ln f^{(k)}(\tau) -\ln f^{(k)}(-\bar\tau) - (k+2) \ln (2\tau_2)
\bigg]\, , \nonumber \\
\een
where
\be{edefk0}
K_0=-{K\over 4\pi} (i)^{-k-2} (-1)^k = {1\over 4\pi\, N^{(k+4)/2}}
\, .
\ee
Note that the degeneracy factor is positive or negative depending on
whether $Q_e\cdot Q_m$ is positive or negative. This is natural
from the point of view of a black hole solution since a classical
black hole is expected to be bosonic (fermionic) for $Q_e\cdot Q_m$
even (odd).

Identifying $|d(Q_e, Q_m)|$ with $e^{S_{stat}(Q_e, Q_m)}$ where
$S_{stat}$ denotes the statistical entropy, we can rewrite
\refb{er7} as
\be{ek1}
e^{S_{stat}(Q_e, Q_m)} = \int{d^2\tau\over \tau_2^2} \, e^{-F(\vec \tau)}\, ,
\ee
where $\vec\tau = (\tau_1, \tau_2)$ or $(\tau, \bar\tau)$ depending on
the basis we choose to use, and
\bea{ek2}
F(\vec\tau) &=& -\Bigg[ {\pi\over 2 \tau_2} \, |Q_e +\tau Q_m|^2
-\ln f^{(k)}(\tau) -\ln f^{(k)}(-\bar\tau) - (k+2) \ln (2\tau_2)
\nonumber \\
&& +\ln\bigg\{K_0 \, \left(
2(k+3) + {\pi\over \tau_2} |Q_e +\tau Q_m|^2\right)
\bigg\}\Bigg] \, ,
\een
 Note that $F(\vec\tau)$ also depends on 
 the charge vectors $Q_e$,
$Q_m$, but we have not explicitly
displayed these in its argument. 
In \refb{ek1}
the integration measure $d^2\tau / (\tau_2)^2$ as well as the integrand
$e^{F(\vtau)}$ 
are manifestly invariant under the $\Gamma_1(N)$
transformation:
\bea{egamman}
&& Q_e \to a Q_e - b Q_m, \quad Q_m \to -c Q_e + d Q_m, \quad
\tau \to {a\tau + b\over c\tau + d}\, , \nonumber \\
&& \quad a,b,c,d\in \ZZZ, \quad ad - bc = 1, \quad a,d=
\hbox{1 mod $N$},
\quad c = \hbox{0 mod $N$}\, .
\een
Thus $S_{stat}(Q_e, Q_m)$ computed from \refb{ek1} is 
invariant under $\Gamma_1(N)$.

We shall now describe a systematic procedure for
evaluating $S_{stat}$ as an expansion in inverse powers of
the charges. For this we
introduce the generating function:
\be{ek3}
e^{W(\vec J)} = \int{d^2\tau\over \tau_2^2} 
\, e^{-F(\vec \tau) + \vec J\cdot
\vec \tau}\, ,
\ee
for a two dmensional vector $\vec J$, and define $\Gamma(\vec u)$ as
the Legendre transform of $W(\vec J)$:
\be{ek4}
\Gamma(\vec u) =  \vec J\cdot \vec u - W(\vec J) \, ,
\qquad u_i = {\p W(\vec J)\over \p J_i}\, .
\ee
It follows from \refb{ek4} that 
\be{ek5}
J_i = {\p \Gamma(\vec u)\over \p u_i}\, .
\ee
As a result if
\be{ek6}
{\p \Gamma(\vec u)\over \p
\vec u_i} = 0 \quad \hbox{at $\vec u 
=\vec u_0$}\, ,
\ee
then it follows from \refb{ek3}-\refb{ek5},
\refb{ek1} that
\be{ek7}
\Gamma(\vec u_0) = -W(\vj=0)= -S_{stat}  \, .
\ee
Thus the computation of $S_{stat}$ can be done by first calculating
$\Gamma(\vec u)$ and then evaluating it at its extremum. 
$\Gamma(\vec u)$ in turn can be calculated by regarding this as a
sum of one particle irreducible (1PI) Feynman diagrams in the zero
dimensional field theory with action $F(\vec\tau)+2\ln\tau_2$.  
Since $S_{stat}$ is given
by the value of the function $-\Gamma(\vu)$ at its extremum, we
can identify $-\Gamma(\vu)$ as the entropy function for the
statistical entropy in analogy with the corresponding result for
black hole entropy\cite{0506177,0508042}.

A convenient method of
calculating $\Gamma(\vu)$ is the so called background field method.
For this we choose some arbitrary base point $\vtau_B$ and define
\be{ek3a}
e^{W_B(\vtau_B,\vec J)} = \int{d^2\eta\over (\tau_{B2}+\eta_2)^2} 
\, e^{-F(\vtau_B+\htau) + \vec J\cdot
\htau}\, ,
\ee
\be{ek4a}
\Gamma_B(\vtau_B, \vc) =  \vec J\cdot \vc - W_B(\vtau_B,\vec J) \, ,
\qquad \chi_i = {\p W_B(\vtau_B,\vec J)\over \p J_i}\, .
\ee
By shifting the integration variable in \refb{ek3a} to $\vtau=\vtau_B
+\htau$ it follows easily that
\be{ek5a}
W_B(\vtau_B,\vec J)=W(\vec J) - \vtau_B\cdot \vj\, ,
\ee
and hence
\be{ek6a}
\Gamma_B(\vtau_B, \vc) = \Gamma(\vtau_B+\vc)\, .
\ee
Thus the computation of $\Gamma(\vu)$ reduces to the computation
of $\Gamma_B(\vu, \vc=0)$. The latter in turn can be computed
as the sum of 1PI vacuum diagrams in the 0-dimensional field theory
with action 
$F(\vu +\htau)+2\ln (u_{2}+\eta_2)$, 
with $\htau$ regarded as fundamental fields, and 
$\vu$ regarded as some fixed background.

While this gives a definition of the statistical
entropy function whose
extremization leads to the statistical entropy, the entropy function
constructed this way is not manifestly duality invariant. This is due
to the fact that since the duality transformation has a non-linear action
on $(\tau_1, \tau_2)$, the generating function $W(\vj)$ defined in 
\refb{ek3} and hence also the effective action $\Gamma(\vu)$ 
defined in \refb{ek4} does not have manifest duality symmetries. Of
course the statistical entropy obtained by extremizing $\Gamma(\vu)$
will be duality invariant since this is given in terms
of the manifestly duality
invariant integral \refb{ek2}. In appendix \ref{s3} we have described a
slightly different construction based on Riemann
normal coordinates which yields a manifestly duality invariant 
statistical entropy function. The result of this analysis is that instead 
of using the function $-\Gamma_B(\vec\tau)$ as the statistical entropy 
function we can use a different manifestly duality invariant function 
$-\wt 
\Gamma_B(\vec\tau)$ as the 
statistical entropy function. $\wt\Gamma_B(\vec\tau_B)$ is defined as
the sum of
1PI vacuum diagrams computed from the action 
\be{einfo1}
-\ln\left( {1\over |\vec\xi|} \sinh{|\vec\xi|} \right) 
- \sum_{n=0}^\infty {1\over n!} (\tau_{B2})^n
\xi_{i_1}\ldots
\xi_{i_n}\, D_{i_1}
\cdots D_{i_n} F(\vtau)\bigg|_{\vtau=\vtau_B}\, ,
\ee
where 
$D_\tau$, $D_{\bar\tau}$ are duality invariant covariant derivatives
defined recursively through the relation:
\bea{einfo3}
D_\tau (D_\tau^m D_{\bar\tau}^n F(\vec\tau))
&=& (\p_\tau - im/\tau_2) (D_\tau^m D_{\bar\tau}^n F(\vec\tau)),
\nonumber \\
D_{\bar\tau} (D_\tau^m D_{\bar\tau}^n F(\vec\tau))
&=& (\p_{\bar\tau} + in/\tau_2)
(D_\tau^m D_{\bar\tau}^n F(\vec\tau))\, ,
\een
for any arbitrary ordering of $D_\tau$ and $D_{\bar\tau}$
in $D_\tau^m D_{\bar\tau}^n F(\vec\tau)$. 
During this computation the components $(\xi,\bar\xi)$ 
or $(\xi_1,\xi_2)$ of
$\vec\xi$ are
to be regarded as the zero dimensional quantum fields and $\vec\tau_B$
is to be taken as a fixed base point.
The result of this
computation expresses $\wt\Gamma_B(\vec\tau_B)$ in terms
of manifestly duality invariant quantity $F(\vec\tau)$ and its
duality invariant covariant derivatives.

It has also been shown in appendix \ref{s3} that explicit evaluation of 
$\wt\Gamma_B(\vec\tau)$ gives
\be{einfo4}
-\wt\Gamma_B(\vec\tau) = {\pi\over 2 \tau_{2}} \, |Q_e +\tau Q_m|^2
- \ln f^{(k)}(\tau) -\ln f^{(k)}(-\bar\tau)
- (k+2) \ln (2\tau_{2}) + \hbox{constant} + \OO(Q^{-2})\, .
\ee
Up to an additive constant this agrees with the black hole entropy
function for CHL models
calculated in \cite{0508042,9906094,0007195} using the quantum
effective action of these theories\cite{9708062,0502126}.

\bigskip

{\bf Acknowledgement}: We would like to thank Dileep Jatkar
for collaboration during initial stages of this work and many
useful discussions. We would also like to thank
A.~Dabholkar, 
D.~Gaiotto, E.~Gava and K.S.~Narain
for useful discussions. J.R.D. would like to thank ASICTP, Trieste for
hospitality during completion of this work. A.S. would like to
thank ASICTP, Trieste and SITP at Stanford University for hospitality
during the completion of the work.
 
 \appendix

\sectiono{Counting Twisted Sector Elementary Heterotic String States in 
CHL 
Models} \label{stwisted}

In this appendix we shall compute
\be{y1}
g(\tau) = \sum_{l_0'} d_{KK}(l_0') e^{2\pi i l_0' \tau} 
\ee
where $d_{KK}(l_0')$ is the degeneracy of the BPS states of a 
Kaluza-Klein monopole associated with the compact direction
$\wt S^1$
carrying momentum $-l_0'$ along $S^1$. 
As discussed in section \ref{sint}, under a duality
transformation $d_{KK}(l_0')$ gets mapped to the degeneracy of
BPS states of a fundamental
string wound once along $S^1$ and carrying momentum
$-l_0'$ along $S^1$ in heterotic string theory on 
$(T^4\times S^1)/\ZZZ_N
\times
\wh S^1$.   Since we have a singly wound string along $S^1$,
after orbifolding it becomes a twisted sector state.
For computing $g(\tau)$ we can now proceed as in 
\cite{0504005}. In heterotic string theory on $T^6$ the
internal momenta belong to the 
Narain lattice embedded in a 28-dimensional Lorentzian
space of signature (6,22)\cite{narain,nsw}.
Let us denote by $g_H$ the generator
of the $\ZZZ_N$ transformation and by  
$V_\parallel$ the subspace of the  28-dimensional momentum space
which transform
non-trivially under $g_H$. 
The dimension of $V_\parallel$ is equal to the number 
of $U(1)$ gauge
fields which are projected out by the orbifolding procedure. This in
turn is given by\cite{0510147}
\be{y1.5}
20 -2k\, , \quad k = {24\over  N+1}-2\, .
\ee
Let $L_0'$ denote the
contribution to the $L_0$ eigenvalue of a state  
from all the left-handed bosonic  oscillators and also
from the components of the momentum along $V_\parallel$, and
$\bar L_0'$ denote
the contribution to the $\bar L_0$ eigenvalue from the
right-handed bosonic and fermionic oscillators.\footnote{All
the components of the right-handed momentum are invariant
under $g_H$ and hence they do not contribute to $L_0'$.}
Then we have
\be{y2}
L_0 - \bar L_0 = L_0' - \bar L_0' - l_0'/N\, ,
\ee
since $-l_0'/N$ is the contribution to $L_0-\bar L_0$ from components
of the internal momentum invariant under $g_H$.
Now level matching condition tells us that $L_0-\bar L_0=0$, whereas
BPS condition requires that we have $\bar L_0'=0$. Thus on these
states
\be{y3}
L_0' = l_0'/N\, .
\ee
This allows us to reexpress $g(\tau)$ as
\be{y4}
g(\tau) = {1\over N} \sum_{s=0}^{N-1}
Tr'_{g_H} \left(g_H^s e^{2\pi i \tau N L_0'}\right)\, ,
\ee
where $Tr'_{g_H}$ denotes trace over states twisted by $g_H$ carrying
all possible momenta along $V_\parallel$ and involving 
arbitrary excitation
of left-moving bosonic oscillators. 
We can simplify this expression by noting that in the twisted
sector of the orbifold theory the $\ZZZ_N$ projection implements
level matching. Since we have already implemented the level matching
condition on the states via eq.\refb{y3}, $g_H$ acts trivially on these
states.
In particular the part of $g_H$ that corresponds to $2\pi/N$ translation
along $S^1$ gives a phase of $e^{-2\pi i l_0'/N}=e^{-2\pi i L_0'}$, and
this cancels the phase coming from the part of $g_H$ that acts on
the twisted bosonic oscillators.
Thus we can rewrite \refb{y4} as
\be{y4a}
g(\tau/N) =  
Tr'_{g_H} \left(  e^{2\pi i \tau L_0'}\right)\, .
\ee
Since the $L_0'$ in the exponent comes from oscillator contribution
and components of momentum along $V_\parallel$, only the part of
$g_H$ twisting which involves the 16 left-moving coordinates and the
coordinates of $T^4$ are relevant for this computation.

Some useful properties of $g(\tau/N)$ are
listed below:
\begin{enumerate}

\item {}From \refb{y4a} we see that $g(\tau/N)$ can be identified
as an appropriate partition function with spin structure $(0,1)$
where the spin structure $(s,r)$ represents
twisting by $g_H^r$ and an insertion of $g_H^s$ 
in the trace.   Since under a modular transformation 
$\tau\to (a\tau+b)/(c\tau+d)$ the spin structure $(s,r)$ gets
transformed to 
\be{y5}
\pmatrix{s\cr r} \to \pmatrix{a & b\cr c & d} \pmatrix{s\cr r}\, ,
\ee
we see that the subgroup of $SL(2,\ZZZ)$ that preserves the set of
spin structures $\pmatrix{0\cr 1}$ mod $N$
is determined by the requirement
\be{y6}
a,b,c,d \in \ZZZ, \quad ad-bc=1, \quad b = \hbox{0 mod $N$}, \quad
d = \hbox{1 mod $N$}\, .
\ee 
This describes the group $\Gamma^1(N)$. 
Thus we expect that $g(\tau/N)$
will transform as a modular form of an appropriate weight under the
group $\Gamma^1(N)$.

\item The weight of the modular form may be determined as follows.
First of all note that $g(\tau/N)$ receives 
contribution from the 24 left-moving
bosonic oscillators. This contributes $-12$ to the 
weight of the modular form. On the other hand $g(\tau/N)$ 
also involves
a sum over a $(20-2k)$ dimensional momentum lattice 
embedded in $V_\parallel$. This gives  a contribution of  $(10-k)$ to
the modular weight. Thus the net modular weight of $g(\tau/N)$ is
\be{y7}
10-k-12 = -k -2=-{24\over N+1}\, .
\ee

\item Since $g(\tau/N)$ receives contribution from
 only bosonic oscillators, each term in the expansion
of $g(\tau/N)$ as a power series in $e^{2\pi i\tau/N}$
has positive coefficients and hence $g(\tau/N)$ does not
vanish in the interior of the upper half plane.

\item For large Im($\tau$), $g(\tau/N)$ behaves as
\be{y8}
g(\tau/N) \simeq 16 e^{-2\pi i\tau/N}\, ,
\ee
where the exponent $-2\pi i\tau/N$ comes from the $L_0'$
eigenvalue of
$-1/N$ associated with the ground state of the 
(twisted) bosonic oscillators\cite{0510147}, and
the factor of 16 counts the 8 Ramond and 8 Neveu-Schwarz sector
states associated with the broken supersymmetry generators. 

\item For small $\tau$ the behaviour of $g(\tau/N)$ is controlled by
the large $Im(\tau)$ behaviour of the partition function with spin
structure (1,0), \i.e. by the trace of $g_H e^{2\pi i L_0'/\tau}$
over the untwisted sector 
states. 
Since the untwisted sector ground state has $L_0'$ eigenvalue
$-1$, this gives:
\be{egbeh}
g(\tau/N) \sim  
e^{2\pi i / \tau } \qquad \hbox{for $\tau\to 0$}\, .
\ee

\end{enumerate}

{}From this it follows that $g(\tau/N)^{-1}$ is a modular form of
$\Gamma^1(N)$ of
weight $(k+2)$ without any pole in the interior of the upper half plane,
and behaves as
\bea{y9}
(g(\tau/N))^{-1} &\simeq& {1\over 16} e^{2\pi i\tau/N}\, , 
\quad \hbox{for
large Im $\tau$}, \nonumber \\
&\sim &
\, e^{-2\pi i /\tau} \quad  \hbox{for
small $\tau$}\, .
\eea
Since $g(\tau/N)^{-1}$ vanishes at the cusps $\tau=n+i\infty$, 
$n\in\ZZZ$, and $\tau=0$, we conclude that it is a cusp form of 
$\Gamma^1(N)$
of weight $(k+2)$.  This leads to a unique choice for
$g(\tau/N)$\cite{mod}:
\be{y10}
g(\tau/N) = {16 \over f^{(k)}(\tau/N)} 
= 16 \eta(\tau/N)^{-k-2} \eta(\tau)^{-k-2}\, .
\ee
This gives
\be{y10a}
g(\tau)=
16\, e^{-2\pi i \tau} \prod_{n=1}^\infty \left\{
(1 - e^{2\pi i n\tau})^{-{24\over N+1}}
(1 - e^{2\pi i nN\tau})^{-{24\over N+1}}\right\}\, .
\ee
For $N=1$ this reproduces the standard result $g(\tau)=
16 \eta(\tau)^{-24}$ for the partition function of half BPS states
in toroidally compactified heterotic string theory\cite{dabh}.
For $N=2$ this result agrees with that of \cite{0502157} obtained
by explicit computation.

 \sectiono{Counting States in the Supersymmetric Field Theory
 with Taub-NUT Target Space}  \label{sb}
 
In this appendix we will analyze
the overall motion of the D1-D5 system in the Taub-NUT 
(Kaluza-Klein monopole) space and
count the number of states of this system which
carry momenta $-l_0$ along $S^1$ and $j_0$ along 
$\wt S^1$. 
The D1-D5-system wrapped on $K3\times S^1$ in flat transverse
space
has four bosonic zero modes
labelling the transverse coordinates and eight fermionic zero modes
associated with the breaking of 
eight out of sixteen supersymmetries of type IIA
string theory on $K3\times S^1$. Thus when the transverse space
is Taub-NUT, we expect
the low energy  dynamics of this system to be
described by a (1+1) dimensional supersymmetric field theory 
with four bosonic and eight fermionic
coordinates, with the bosonic coordinates taking value in the 
Taub-NUT 
target space.  Since the world-sheet coordinate
$\sigma$ of this field theory
is identified with the coordinate along $S^1$, and
since $S^1/\ZZZ_N$ has periodicity $2\pi/N$, the natural unit
of momentum along $\sigma$ is $N$.  
As a result a  state of the D-brane system carrying momentum 
$-l_0$ 
corresponds to a state in this field theory 
with $L_0 -\bar L_0= l_0/N$. 

We begin our analysis of this field theory by writing down the
metric of the Taub-NUT space:
\be{tnutgeom}
ds^2 = \left(1+\frac{R}{r}\right) 
\left( dr^2 +  r^2 ( d\theta^2 + \sin^2 \theta d\phi^2) \right)
+ R^2\left( 1 + \frac{R}{r}
\right)^{-1} ( 2\, d\psi + \cos\theta d\phi)^2
\ee
with the identifications:
\be{eident}
(\theta,\phi,\psi) \equiv (2\pi -\theta,\phi+\pi, \psi+{\pi\over 2})
\equiv (\theta,\phi+2\pi,\psi+\pi)\equiv (\theta,\phi,\psi+2\pi)\, .
\ee
Close to the origin the metric reduces to that of flat space $R^4$ 
written in terms of Euler angles $\theta, \phi, \psi$ and $r$,
while for large $r$ it is that of $R^3\times \wt S^1$, with
$\wt S^1$ parametrized by the angular coordinate $\psi$.  
For later
use it will be useful to introduce the cartesian coordinates:
\bea{ecart}
&& x^1 = 2\sqrt r \cos{\theta\over 2}\cos\left(
\psi+{\phi\over 2} \right), \qquad x^2 =  
2\sqrt r \cos{\theta\over 2}\sin\left(
\psi+ {\phi\over 2} \right), \nonumber \\ &&
 x^3 =  2\sqrt r \sin{\theta\over 2}\cos\left(\psi -
{\phi\over 2}\right), 
\qquad x^4 =  2\sqrt r \sin{\theta\over 2}\sin\left(\psi -
{\phi\over 2}\right)\, .
\eea

The metric \refb{tnutgeom} has a global $U(1)_L\times SU(2)_R$
symmetry. The $SU(2)_R$ symmetry refers to the usual rotation
group of three dimensional space and will not be of interest
to us. The $U(1)_L$ symmetry acts as
\be{eu12}
\psi\to\psi+\epsilon\, ,
\ee
with no action on any of the other coordinates. From the point of view 
of an asymptotic observer 
this is just a translation along the compact circle $\wt S^1$
parametrized by $\psi$, and the
corresponding conserved charge is the quantum number $j_0$.
On the other hand using \refb{ecart} 
we see that near the origin
the $\psi$ translation acts as simultaneous rotation
along the 1-2 and 3-4 planes. Thus near the origin
the contribution to the $j_0$ charge can be identified
as the sum of the angular momentum in the 1-2 and 3-4 planes.

Let us now analyze the transformation laws of the fermion
fields under $U(1)_L$ transformation. Since the fermions transform under
the tangent space group, we need to find the compensating
tangent space rotation
that accompanies the global $U(1)_L$ transformation.
Let $SO(4)^T=SU(2)^T_L\times SU(2)^T_R$ be the tangent space
rotation group of the Taub-NUT space, -- 
this needs to be distinguished
from the global symmetry group described earlier. 
The Taub-NUT
space is known to have $SU(2)$ holonomy. We shall choose our 
convention such that the holonomy is in the $SU(2)^T_L$.
To see what this means 
we use the fact that
for a suitable choice of the tangent space basis vectors $(e^0,e^1,e^2,
e^3)$ the spin
connection $\omega^{ab}=\omega^{ab}_\mu dx^\mu$ 
associated with the
metric \refb{tnutgeom} is given by\cite{eguchi}
\bea{espin}
&&\omega^{01} = -\left( r + {R\over 2}\right) 
\left( r + {R }\right)^{-1}\, \sigma_x\, , \qquad
\omega^{02} = -\left( r + {R\over 2}\right) 
\left( r + R \right)^{-1}\, \sigma_y\, , \nonumber \\ 
&&
\omega^{03} = -{R^2\over 2 } 
\left( r + {R}\right)^{-2}\, \sigma_z\, , 
\qquad
\omega^{23} = -{R\over 2}  
\left( r + {R }\right)^{-1}\, \sigma_x\, , \nonumber \\
&& \omega^{31} = -{R\over 2} (r+R)^{-1}
\, \sigma_y, \qquad
\omega^{12} = \left({R^2\over 2} (r + R)^{-2} -1\right) \sigma_z
\,,
\eea
where $\sigma_x$, $\sigma_y$ and $\sigma_z$ are a set of one-forms
\bea{eoneform}
\sigma_x &=& 
\cos\left({2\psi }\right) d\theta + \sin\left({2\psi }\right)
 \sin\theta d\phi\, , \nonumber \\
 \sigma_y &=& 
-\sin\left({2\psi }\right) d\theta + \cos\left({2\psi }\right)
 \sin\theta d\phi\, , \nonumber \\
\sigma_z &=& \cos\theta d\phi +2 d\psi\, .
\eea
Thus near the origin $r=0$ we have
\be{orig}
\omega^{01}=\omega^{23} = -{1\over 2} \sigma_x, 
\qquad \omega^{02}=\omega^{31} = -{1\over 2}\sigma_y,
\qquad \omega^{03}=\omega^{12} = -{1\over 2} \sigma_z
\, .
\ee
If $\Sigma^{ab}$ denotes the generator of the tangent
space group  then we see from
\refb{orig} that near the origin the holonomy group is
generated by the combinations $\Sigma^{01}+\Sigma^{23}$,
$\Sigma^{02}+\Sigma^{31}$ and $\Sigma^{03}+\Sigma^{12}$.
These generate an SU(2) group, -- by our convention this
describes the tangent space $SU(2)_L^T$ group near the origin. The
precise relation between the $\Sigma^{ab}$'s and the generators
$T^1$, $T^2$, $T^3$ of $SU(2)_L^T$ are
\be{eprecise}
T^1={1\over 2}(\Sigma^{01}+\Sigma^{23}), \qquad
T^2={1\over 2}(\Sigma^{02}+\Sigma^{31}), \qquad
T^3={1\over 2}(\Sigma^{03}+\Sigma^{12})\, ,
\ee
so that we have
\be{eomegaab}
\omega^{ab}\Sigma^{ab} = -\sigma_x T^1 -\sigma_y T^2 -
\sigma_z T^3\, .
\ee
Using \refb{eoneform} we can express this as
\be{eomega2}
\omega^{ab}\Sigma^{ab} =- i U_0\, d \, U_0^{-1}\, ,
\ee
where
\be{eomega3}
U_0 = e^{2i\psi T_3} e^{i\theta T_1} e^{i\phi T_3}\, .
\ee

Now let us consider the effect of a global $U(1)_L$ rotation
$\psi\to\psi+\epsilon$.
Using \refb{eomega2}, \refb{eomega3} we see that this induces a
transformation
\be{eomtrs}
\omega^{ab}\Sigma^{ab}\to e^{2i\epsilon T_3} 
\omega^{ab}\Sigma^{ab} e^{-2i\epsilon T_3}\, .
\ee
Since $T_3$ is a generator of $SU(2)_L^T$ we see that in order to
preserve the action a global $U(1)_L$ rotation must be accompanied
by a compensating $SU(2)_L^T$ transformation. In particular 
the fermion fields
$\chi$
must transform as
\be{eomtrs2}
\chi\to e^{2i \epsilon T_3} \chi\, ,
\ee
where $T_3$ is to be taken in the representation of $SU(2)_L^T$
in which the fermions transform.

Now we know that the fermions transform in the $(1,2)+(2,1)$
representation
of the tangent space $SU(2)_L^T\times SU(2)_R^T$ group.
Thus half of the fermions are neutral under $SU(2)_L^T$
and hence also under the global $U(1)_L$. On the other hand
since the holonomy is in $SU(2)_L^T$, there is no coupling
of these fermions to the spin connection and they behave as
free fermions. 
The other half of the fermions, carrying non-trivial
$SU(2)_L^T$ quantum numbers, are interacting and
do transform under the
global $U(1)_L$ group.
Furthermore, since type IIB string theory on K3
is a chiral theory, the world-sheet chirality of these fermions is
correlated with the chirality under the tangent space rotation group.
In particular, the $SU(2)_L^T$ singlet free fermions are left-moving
on the D1-D5-brane world-sheet,  and the $SU(2)_R^T$ singlet
interacting fermions are 
right-moving on the world-sheet.\footnote{This can be seen as follows.
Since the bosons are interacting, the free fermions 
have no bosonic
superpartner, and hence they do not transform under the 
unbroken supersymmetry of the
D1-D5 system in K3$\times$Taub-NUT space. Thus they must
be left-moving. On the other hand the $SU(2)_R^T$ singlet interacting
fermions are the superpartners of the interacting bosonic fields
and hence must be right-moving on the world-sheet.}

This shows that the resulting (1+1) dimensional 
world-volume theory on the D1-D5 system
is described by a set of four free left-moving
$U(1)_L$ invariant fermion fields, together with an
interacting theory of four bosons and four right-moving
$U(1)_L$ non-invariant fermions. Let us first
calculate the contribution to the partition function due to the free
left-moving
fermions. Since these fermons do not carry any $j_0$ charge, their
contribution is given by:
\bea{ezfree}
Z_{{\rm free}}(\hat \rho) \equiv {\rm Tr}_{{\rm free \, 
left-moving\,
fermions}}
( (-1)^{F+\bar F} e^{2\pi i \hat \rho l_0 
+ 2\pi i \hat v j_0} ) , \cr
=4\, \prod_{n=1}^\infty( 1 - e^{2\pi i n N \hat \rho})^4 \, ,
\eea
taking into account the identification $l_0=(L_0-\bar L_0)/N$.
The factor of $4$ comes from the quantization of the free fermion
zero modes. The latter in turn can be interpreted as due to the four
broken supersymmetries of the D1-D5-system 
on $K3\times$Taub-NUT space.

Now we turn to the interacting part of the theory. Since we are
computing an index we can assume that it does not depend on
continuously varying parameters. Let us take the size $R$ of the
Taub-NUT space to be large so that the metric is almost flat and
in a local region of the Taub-NUT space the world-volume
theory of the D1-D5 system is almost
free. In this case we should be able to compute the contribution to the
non-zero mode bosonic and fermonic oscillators by placing the
D1-D5 system at any point in Taub-NUT space, -- say at the
origin. The contribution from the zero modes however is sensitive
to the global geometry of the Taub-NUT space and should be
computed separately.
 
Since supersymmetry acts on the right-moving bosons and fermions,
in order to get a BPS state the right-moving bosonic and fermionic
oscillators must be in their ground state. Thus as far as the
contribution due to the non-zero mode oscillators are
concerned, we only need to
examine the effect of  left-moving
bosonic oscillators carrying
$L_0=l_0/N$, $\bar L_0=0$
and angular momentum $j_0$.  {}From \refb{ecart}
we see that  a translation of $\psi$ induces simultaneous rotation
in the $x^1-x^2$ and $x^3-x^4$ plane. Hence we need to
compute the partition function of free left-moving
bosons carrying total angular
momentum $j_0$ in the $x^1-x^2$ and $x^3-x^4$ plane.
This is easily done using the result of \cite{9405117}. The answer
is  
\bea{rusosus}
Z_{{\rm osc}}(\hat \rho, \hat v) &\equiv&  {\rm Tr}_{\rm{oscillators}} 
( (-1)^{F+\bar F}
e^{2\pi i \hat \rho l_0 + 2\pi i \hat v j_0} ) \cr
&=& 
\prod_{n=1}^\infty \frac{1}
{(1- e^{2\pi i n N \hat \rho + 2\pi i \hat v})^2
(1- e^{2\pi i n N \hat \rho - 2\pi i \hat v})^2 }\, .
\eea
The result can be understood as follows. The 
bosonic oscillators in the $1-2$ plane and the $3-4$ plane can be
split into complex conjuate pairs which carry opposite units of 
angular momentum. 
Therefore two of the 
bosonic oscillators carry $+1$ unit of angular momentum and
the other two 
carry $-1$ unit of angular momentum. The partition function 
weighted with the angular momentum is then given by \eq{rusosus}.

Finally we turn to the contribution $Z_{{\rm z}}$ to the
partition function from the bosonic and
fermionic zero modes of the interacting part of the theory. 
Since the intercting theory has four bosonic and four fermionic
fields, the dynamics is that
of a
superparticle with four bosonic and four fermionic coordinates
moving in Taub-NUT space. 
Under the holonomy group
$SU(2)^T_L$ 
of the Taub-NUT space both the bosons and the fermions 
transform in a pair of spinor representations. This system is
described by
an $N=4$ supersymmetric quantum mechanics. 
Thus in order to look for
BPS states of the D1-D5 system we need to look for supersymmetric
ground states of this  quantum mechanics.

So far we have been working at a special point in the moduli
space of the CHL string theory where the circles $S^1$ and $\wt S^1$
are orthonormal in the asymptotic geometry. In this case the BPS mass
of the D1-D5-Kaluza-Klein monopole system is
equal to the sum of the BPS masses of the D1-D5 system and
the Kaluza-Klein monopole. As a result there is no potential
term in the D1-D5 world-volume action and analysis of bound 
states is difficult.
But this is not a generic situation. Once we
switch on a component of the metric that mixes $S^1$
and $\wt S^1$ the BPS mass of the D1-D5-Kaluza-Klein monopole
system is
less than the sum of the BPS masses of the D1-D5 system and
the Kaluza-Klein
monopole. In this case
the D1-D5 system in the presence of the monopole has
a potential and the system is easier to analyze. On the
other hand the analysis of the dynamics of non-zero modes will not
be
affected by this modification since we are computing an index, 
-- hence our results for $Z_{\rm free}$ and
$Z_{\rm osc}$ should remain unchanged.

The mixing between $S^1$ and $\wt S^1$ can be achieved
by replacing the $d\psi$ term
in the expression for the metric given in \refb{tnutgeom}
by $d\psi + a d y$ where $y$ is the coordinate along $S^1$ and $a$
is a small deformation parameter. This clearly remains a solution
of the equations of motion but gives an $r$ dependent contribution
to the $yy$ component of the metric:
\be{egyy}
\Delta g_{yy} = 4\, a^2\,  R^2 \left( 1 +{R\over r}\right)^{-1}\, .
\ee
As a result the tension of the D1-D5 system, being proportional to
$\sqrt{g_{yy}}$, acquires an $r$-dependent contribution proportional
to
\be{egyy1}
a^2 R^2 \left( 1 +{R\over r}\right)^{-1}
\ee
to first order in $a^2$. Supersymmetrization of this term gives rise to
other fermionic terms.

Thus we have to analyze the dynamics of a superparticle with $\NN=4$
supersymmetry moving in 
Taub-NUT space under a potential proportional to \refb{egyy1}.
This is precisely the problem analyzed in 
\cite{9912082,pope} in a different
context. The result of this analysis can be summarized as follows.
Depending on the sign of the deformation parameter $a$ we have
supersymmetric bound states for $j_0>0$ or $j_0<0$, where $j_0$
is the momentum conjugate to the coordinate $\psi$. In the weak
coupling limit the
number of bound states for a given value of $j_0$ is given by
$|j_0|$. If for
definiteness we choose the sign of $a$ such that we get bound states
for positive $j_0$, then this gives the zero mode partition function
\be{ezfin}
Z_{\rm z}(\hat v) 
= Tr_{\rm zero\, modes} ((-1)^{F+\bar F} 
e^{2\pi i \hat v j_0})
= \sum_{j_0=1}^\infty j_0\, e^{2\pi i \hat v j_0} 
={e^{2\pi i\hat v}\over (1 - e^{2\pi i\hat v})^2}\, .
\ee
Since this is invariant under $\hat v\to -\hat v$ we shall get the
same answer if we had chosen to work with the opposite sign
of $a$ that produces bound states with negative $j_0$.

Finally putting all the ingredients together the partition 
function of states associated with the centre of mass motion 
of the D1-D5 system in Taub-NUT space is given by
\bea{fullparttn}
&& \sum_{l_0, j_0} d_{CM} (l_0, j_0) 
e^{2\pi i l_0 \hat \rho + 2\pi i j_0 \hat v} 
= Z_{{\rm free}}(\hat\rho) 
Z_{\rm osc}(\hat\rho, \hat v)  
Z_{{\rm z}} (\hat v)  
\\ \nonumber
&& \qquad = 4 e^{-2\pi i \hat v} ( 1- e^{-2\pi i \hat v}) ^{-2}
\\ \nonumber 
&& \qquad \times \prod_{n=1}^\infty \{
( 1- e^{2\pi i nN \hat \rho})^4  
( 1 - e^{2\pi i nN \hat \rho + 2\pi i \hat v} )^{-2} 
( 1 - e^{2\pi i nN \hat \rho - 2\pi i \hat v} )^{-2} 
\}.
\eea

 \sectiono{Proof of $\wt\Phi_k(T,U,V) = \wt\Phi_k(NU,T/N, V)$} 
\label{sproof}

The Siegel modular form which appears in the expression for the
dyon spectrum of the $\ZZZ_N$ CHL model 
is given by\cite{0602254} 
\bea{es05}
 && \wt\Phi_{k}(U,T,V) = -(i\sqrt N)^{-k-2}
 \exp\left(2\pi i \left( {1\over N}\,
T+ U + V\right) \right) \nonumber \\
&& \qquad
\prod_{r=0}^{N-1}
\prod_{l,b\in \zzz, k'\in \zzz+{r\over N}\atop
k',l,b>0}\Bigg\{ 
 1 - \exp( 2\pi i( k'T + l U + bV)) \Bigg\}^{{1\over 2}\sum_{s=0}^{N-1}
e^{-2\pi ils/N}\, c^{(r,s)}(4lk' -b^2)
  }  \nonumber \\
  && \qquad \prod_{r=0}^{N-1}
\prod_{l,b\in \zzz, k'\in \zzz-{r\over N}\atop
k',l,b>0}\Bigg\{ 
 1 - \exp( 2\pi i( k'T + l U + bV)) \Bigg\}^{{1\over 2}\sum_{s=0}^{N-1}
e^{2\pi ils/N}\, c^{(r,s)}(4lk' -b^2)
  }  \nonumber \\
 \eea
 where $(k', l , b)>0$ means $k'> 0 , l\geq 0, b \in\ZZZ$ or $ k' =0,
l>0, b \in\ZZZ$ or $k' =0, l =0, b <0$.
The coefficients $c^{(r,s)}$ have been defined in
\refb{fifth}-\refb{esi4b}. {}From this definition it follows
that 
 $c^{(r,s)}(4n - b^2) = c^{(-r,-s)}(4n - b^2)$. Hence the two products
in \refb{es05} give identical results and we may rewrite this as 
 \bea{es05a}
 \wt\Phi_{k}(U,T,V) &=&  -(i\sqrt N)^{-k-2}
\, \exp\left(2\pi i \left( {1\over N}\,
T+ U + V\right) \right) \nonumber \\
&&
\prod_{r=0}^{N-1}
\prod_{l,b\in \zzz, k'\in \zzz+{r\over N}\atop
k',l,b>0}\Bigg\{ 
 1 - \exp( 2\pi i( k'T + l U + bV)) \Bigg\}^{\sum_{s=0}^{N-1}
e^{-2\pi ils/N}\, c^{(r,s)}(4lk' -b^2)
  }  \nonumber \\
\eea
For $N=1$ this reduces to the familiar weight 10 cusp form of the
modular group of genus 
two Riemann surfaces\cite{borcherds,9504006,9512046}.

 If we define $j = N k'$ in \refb{es05a}, then in the 
 above product $r=j$ mod $N$, 
 and we
 get
 \bea{es05b}
 \wt\Phi_{k}(U,T,V) &=& -(i\sqrt N)^{-k-2}
\, \exp\left(2\pi i \left( {1\over N}\,
T+ U + V\right) \right) \nonumber \\
&&
\prod_{l,b\in \zzz, j\in \zzz\atop
j,l,b>0}\Bigg\{ 
 1 - \exp( 2\pi i( jT/N + l U + bV)) \Bigg\}^{ \sum_{s=0}^{N-1}
e^{-2\pi ils/N}\, c^{(j,s)}(4lj/N -b^2)
  }  \nonumber \\
\eea
Substituting $U\to T/N$, $T\to UN$ in this product and  exchanging
the dummy indices $j$ and $l$ we get:
\bea{es05c}
&& \wt\Phi_{k}(NU,T/N,V) = -(i\sqrt N)^{-k-2}
\, \exp\left(2\pi i \left( {1\over N}\,
T+ U + V\right) \right) \nonumber \\
&&
\qquad \prod_{l,b\in \zzz, j\in \zzz\atop
j,l,b>0}\Bigg\{ 
 1 - \exp( 2\pi i( jT/N + l U + bV)) \Bigg\}^{ \sum_{s=0}^{N-1}
e^{-2\pi ijs/N}\, c^{(l,s)}(4lj/N -b^2)
  }  \nonumber \\
\eea
Now using \refb{fifth} one can check explicitly that\footnote{For
this we need to use the relation $\sum_{k=0}^{N-1} E_N
\left({\tau+k\over N}\right) = N E_N(\tau)$. This follows from the
Fourier expansion of $E_N(\tau)$ given in \refb{third}.}
\be{es05d}
\sum_{s=0}^{N-1}
e^{-2\pi ils/N}\, F^{(j,s)}(\tau, z)
= \sum_{s=0}^{N-1}
e^{-2\pi ijs/N}\, F^{(l,s)}(\tau, z)\, ,
\ee
and hence from  \refb{esi4b}
\be{es05e}
\sum_{s=0}^{N-1}
e^{-2\pi ils/N}\, c^{(j,s)}(4n - b^2)
= \sum_{s=0}^{N-1}
e^{-2\pi ijs/N}\, c^{(l,s)}(4n-b^2)\, .
\ee
Eq.\refb{es05e} shows 
that the right hand sides of \refb{es05b}
and \refb{es05c} are identical. Hence 
\be{es05f}
\wt\Phi_k(T,U,V) = \wt\Phi_k(NU,T/N, V)\, .
\ee

\sectiono{The 
 D1-D5 System and SCFT with Target Space $S^P K3/\ZZZ_N$} 
\label{ssym}

In this appendix we evaluate the degeneracy of the
low lying states of the
D1-D5  system described in sections \ref{sseq} and
\ref{sint}
for $Q_5 \geq 1$. Before the $\ZZZ_N$ orbifolding
the  dynamics of the relative motion
of
$Q_5$ D5-branes wrapped  on $K3\times S^1$ and $Q_1$ 
D1-branes wrapped on $S^1$ is captured by the ${\cal N} =(4,4)$
superconformal $\sigma$-model with the symmetric product of 
$Q_5(Q_1 -Q_5) +1$ copies of $K3$ as the target 
space\cite{9512078}.   We shall denote this target space by 
$S^P K3\equiv
(K3)^P/S_P$, where $S_P$ refers to the permutation group of $P$
elements. The world-sheet coordinate $\sigma$ of this conformal
field theory is identified with the coordinate along $S^1$.
We shall first review various aspects of the superconformal field
theory with target space $ (K3)^P/S_P$, and then discuss the effect
of the $\ZZZ_N$ projection that is required to describe a
D1-D5-brane configuration on the CHL orbifold.

 Let $g$ be an element of  $S_P$ and $[g]$ denote the conjugacy class
 of $g$. 
Then the Hilbert space of the SCFT with target space $(K3)^P/S_P$
decomposes into a direct sum of twisted sectors labelled 
by the conjugacy classes of $S_P$:
\be{dechilb}
{\cal H}= \oplus_{[g]} {\cal H}_{g}^{({\cal C}_g)} 
\ee
where $\CC_g$ denotes the centralizer of $[g]$ and
${\cal H}_{g}^{({\cal C}_{g}) }$ refers to the
Hilbert space in the $g$ twisted sector projected by $\CC_g$.
The conjugacy classes of $S_P$ may be labelled as
\be{decompg}
[g] = (1)^{P_1} (2)^{P_2} \cdots (s)^{P_s} 
\ee
where $(w)$ denotes cyclic permutation of $w$ elements and
$P_w$ is the number of copies of $(w)$ in $g$. Thus these
conjugacy classes 
are characterized by partitions $P_w$ of $P$ such that
\be{consconj}
\sum_w wP_w = P\, .
\ee
The 
centralizer $\CC_g$ of the conjugacy class $[g]$ given in \refb{decompg}
is given by
\be{central}
{\cal C}_{g} =
S_{P_1} \times
 (S_{P_2} \times \ZZZ_2^{P_2})   \times
\cdots\times
(S_{P_s} \times \ZZZ_s^{P_s})  \, .
\ee
If we denote by $\HH_w$ 
the Hilbert space of states twisted by the
generator $\omega$ of the $\ZZZ_w$ group of
cyclic permutation of $w$ elements, and projected 
by the same $\ZZZ_w$ group, then \refb{central}
shows that for the conjugacy class $[g]$
given in \refb{decompg}
\be{symmtenpro}
{\cal H}_{g}^{({\cal C}_{g}) } 
=
\otimes_{w>0}
S^{P_w}{\cal H}_{w} 
\ee

Consider first the Hilbert space ${\cal H}_{w}$.  
This twisted 
sector is represented by the Hilbert space of the sigma model
of $w$ coordinate fields $X_i(\sigma) \in K3$ with the 
cyclic boundary condition
\be{cycbc}
X_i(\sigma + 2\pi) = \omega X_i(\sigma) = X_{i+1}(\sigma), \quad
i\in (1, \ldots , w)\, ,
\ee
where $\omega$ acts by $\omega: X_i \rightarrow X_{i+1}$. 
Therefore the $w$ coordinate fields can be glued together as a single
field  but in the interval $0\leq \sigma\leq 2\pi w$, moving
in the target space K3. Thus we now have a string of length
$2\pi w$, -- commonly known as the long string, -- moving in K3.
The natural unit of momentum on this string is $1/w$, and hence
if a state of this string carries physical momentum $-l$ along $S^1$,
it corresponds to $L_0-\bar L_0$ 
eigenvalue $lw$.

At this stage it is worth pointing out that whereas for $Q_5=1$ the
quantum number $w$ can be identified with the winding charge of
the D-string, this is not so for $Q_5>1$. Thus we should not regard
the long string as a D-string, -- rather it provides 
some effective description
of the dynamics.  

Once we know the spectrum of $\HH_w$, 
the full spectrum of the CFT of the D1-D5 system is obtained by
taking the direct product of the spectrum of $\HH_w$'s
and then carrying out appropriate symmetrization described in
\refb{symmtenpro}.

We now turn to the effect of the $\ZZZ_N$ projection that is required
in order to get a state of the D1-D5 system in the $\ZZZ_N$ CHL model.
For this we need to pick a $\ZZZ_N$ invariant state
from each of the 
$\HH_w$.
Thus we need to first study the effect of the $\ZZZ_N$ projection in
each $\HH_w$ 
and then take the direct product of these $\ZZZ_N$ invariant
subspaces followed
by appropriate symmetrization. 
In particular if we denote by $\HH_w'$ the subspace of $\ZZZ_N$ 
invariant states in $\HH_w$, then the full $\ZZZ_N$ invariant 
Hilbert space of the D1-D5 system will be given by
\be{etar10}
\otimes_{w>0}
S^{P_w}{\cal H}'_{w} \, .
\ee
The $\ZZZ_N$ projection on a single
long string is implemented as follows. Since the long string has length
$2\pi w$, we need to impose periodic boundary condition on this
string with period $2\pi w/N$. Since under a translation by $2\pi w/N$
on the string the physical coordinate along $S^1$ gets shifted by
$2\pi r/N$ where $r=w$ mod $N$,
under such a shift the coordinates of the string along K3 
must be transformed by $\wt g^r$  where $\wt g$
is the generator of the $\ZZZ_N$ action on K3. Thus if $\wt\ZZZ_N$
denotes the group generated by $\wt g$, we can view the
long string as a closed string of length $2\pi w/N$ with target
space $K3/\wt\ZZZ_N$, 
belonging to a sector
of this SCFT twisted by $\wt g^r$. Since the natural unit of momentum
along the string is now $N/w$, a physical momentum $-l$ along $S^1$
would correspond to $L_0-\bar L_0$ 
eigenvalue of $lw/N$ in the SCFT describing
the dynamics of the long string. 

Now from \refb{esi5} we know that for an SCFT with target
space $K3/\wt\ZZZ_N$ 
\be{etar1}
{1\over N}\, Tr_{RR;\wt g^r} (\wt g^s (-1)^{F+\bar F}
\delta_{NL_0,lw} \delta_{\JJ, j})
= c^{(r,s)}(4lw/N - j^2)\, .
\ee
Here $\JJ$ is the angular momentum operator.
Also the projection operator for $\ZZZ_N$ invariant states with
physical momentum $-l$ along $S^1$ is:
\be{etar2}
{1\over N} \sum_s e^{-2\pi i l s} \wt g^s\, .
\ee
Hence the total number of bosonic minus fermionic states
in the single
long string Hilbert space, carrying
momentum $-l$ along $S^1$  
and angular momentum $j$  is given by:
\be{etar3}
{1\over N} \sum_s e^{-2\pi i l s} 
Tr_{RR;\wt g^r} (\wt g^s (-1)^{F+\bar F}
\delta_{NL_0,lw} \delta_{\JJ, j}) = \sum_s e^{-2\pi i l s} 
c^{(r,s)}(4lw/N - j^2)\equiv n(w,l,j)\, .
\ee

According to
\eq{etar10}  the next step is the evaluation
of the partition function for the symmetrized 
tensor products of 
the Hilbert spaces 
${\cal H}_w'$. For this we use the
following formula from \cite{9608096}. 
If $d_{sym}(P_w,w,L,J)$ denotes the number of bosonic minus
fermionic states in $S^{P_w} {\cal H}_w'$ carrying total momentum
$-L$ along $S^1$ and total angular momentum $J$, then
\be{etar4}
\sum_{P_w=0}^\infty \sum_{L,J}
d_{sym}(P_w,
w,L,J) e^{2\pi i L\wh \rho
+2\pi i J\wh v + 2\pi i y P_w}
= \prod_{l, j\in \zzz\atop l\ge 0} \left( 1 - e^{2\pi i y + 2\pi i l\wh\rho
+ 2\pi i j \wh v}\right)^{-n(w,l,j)}\, .
\ee
The proof of this given in \cite{9608096} holds 
provided that the coeficients $n(w,l,j)$ are integers.
Therefore to apply this formula for the Hilbert space
${\cal H}_w'$ we need  
\be{intecond}
n(w,l,j) = \sum_{s=0}^{N-1} c^{(r,s)}(4lw/N - j^2) 
e^{ {-2\pi i l s}/{N}}
\ee
to be an integer. This can be explicity verified using the 
formulae given in \eq{fifth}-\eq{third}.

Using the  identity in \eq{etar4} we can evaluate the generating function
for the bosonic minus fermionic states for the relative
dynamics of the D1-D5 system. 
Eq.\refb{etar10} shows that all we
need to do is to take the product over $w$ of the right hand side of
\refb{etar4}. More specifically, if $h'(P,n,J)$ denotes the total number
of bosonic minus fermionic states carrying total string length $2\pi
P/N =2\pi \sum_{w>0} w P_w/N$
(counting a single string with quantum number $w$ to have length
$2\pi w/N$), total momentum $-n$ along $S^1$ and total angular
momentum $J$, then we have
\bea{d1d5genfn1}
&\,&\sum_{P, n, J} h'( P, n,J) e^{2\pi i (\hat\rho n + \hat \sigma( P -1)/N +
\hat v J )}  \nonumber \\
&\,& = e^{-2\pi i \hat \sigma /N} \, \prod_{w\in \zzz\atop w>0}
\prod_{l, j\in \zzz\atop l\ge 0} \left( 1 - e^{2\pi i \wh\sigma w /N
+ 2\pi i l\wh\rho
+ 2\pi i j \wh v}\right)^{-n(w,l,j)} \nonumber \\
&\,& =  
e^{-2\pi i \hat \sigma /N} 
\prod_{r=0}^{N-1}\prod_{ p'\in\zzz+ \frac{r}{N}, l, j\in\zzz\atop
p'>0, l\ge 0} 
( 1- e^{2\pi i (p'\hat\sigma + l \hat \rho + j\hat v ) })
^{- \sum_{s=0}^{N-1} 
e^{-2\pi i s l/N} c^{(r,s)}( 4 p' l - j^2) } \, .
\nonumber \\
\eea
We have multiplied $e^{-2\pi i \hat \sigma/N}$
on both sides of the above equation to absorb the shift of unity in 
the value of $P$.  Also in arriving at the last expression in
\refb{d1d5genfn1} we have defined $p'=w/N$.
Physically the quantum number $P$ corresponds
to  $P= Q_5(Q_1-Q_5) +1$.

Let us now evaluate the full degeneracy 
$h( P, n,J)$ of the D1-D5 system and the 
Taub-NUT space, counting each supermultiplet as one state.  
To do this we need to put in the contribution from
the excitation modes of the Kaluza-Klein monopole and the center
of mass motion of the D1-D5 system.
These have already been
computed in appendices \ref{stwisted}, \ref{sb}. Putting together all
the results we get
\bea{d1d5fullgen1}
f(\hat\rho, \hat \sigma, \hat v) &\equiv&
\sum_{P,n,J} h( P, n,J) e^{2\pi i (\hat\rho n + \hat \sigma( P -1)/N +
\hat v J } \cr
&=& \frac{1}{64} 
e^{-2\pi i \hat \sigma /N} 
\prod_{r=0}^{N-1}\prod_{ p'\in\zzz+ \frac{r}{N}, l, j\in\zzz
\atop p'>0, l\ge 0} 
( 1- e^{2\pi i (p'\hat\sigma + l \hat \rho + j\hat v ) })
^{- \sum_{s=0}^{N-1} 
e^{-2\pi i s l/N} c^{(r,s)}( 4 p' l - j^2) } \cr
&\;& \left( \sum_{l_0, j_0} d_{CM} ( l_0, j_0) e^{2\pi i ( l_0\hat \rho
+ j_0\hat v)}  \right)
\left(
\sum_{l_0'} d_{KK} (l_0') e^{2\pi i l_0'\hat \rho} 
\right)
\eea
The factor of $1/64$ in this expression accounts for the fact that each
supermultiplet contains 64 states.
Using \eq{etwo}, \eq{eone}  
one can reduce the above equation  to 
\be{finalfd1}
f(\hat \rho, \hat \sigma, \hat v ) =
e^{-2\pi i (\hat \sigma/N + \hat \rho+ \hat v )}
\prod_{r=0}^{N-1}\prod_{ p'= \zzz+ \frac{r}{N}, l\in\zzz, j
\atop p'\geq 0, l\geq 0} 
( 1- e^{2\pi i (p'\hat\sigma + l \hat \rho + j\hat v  })
^{- \sum_{s=0}^{N-1} 
e^{ {-2\pi i s l}/{N}} c^{(r,s)}( 4 p' l - j^2) } 
\ee
where the product over $j$ runs over integer values for $(p',l)\ne (0,0)$
and runs over only positive integers (or negative integers) when
$p'=l=0$.
Note that  the $p'=0$ terms in the above equation arise
from the terms involving $d_{CM}(l_0, j_0)$ and $d_{KK} (l_0')$.
Now comparing the right hand side of this equation with 
\eq{es05a} we can write 
\be{finphirel}
f(\hat\rho, \hat\sigma,\hat v) =-
\frac{(i\sqrt N)^{-k-2}}{\Phi_k(\hat\rho, \hat\sigma, \hat v)}
\ee
From this point onward the analyis of the degeneracy of the D1-D5 system
in Taub-NUT space proceeds as in section 3 with $Q_m^2 
= 2 Q_5(Q_1-Q_5)$.
We have thus extended the analysis of section \ref{sint} to the
$Q_5\geq 1$ case.

\sectiono{Zeroes and Poles of $\wt\Phi_k$} \label{szero}

As has been shown in \cite{0602254}
the Siegel modular form $\wt\Phi_k(\wrh ,\ws,\wv )$ 
given in \refb{es05a}
satisfies the relation:
\bea{eu1}
&& -2\ln \wt\Phi_k - 2\ln \bar{\wt\Phi}_k -2 \, k\, \ln \det Im
\Omega + \hbox{constant}
\nonumber \\
 &=& \sum_{r,s=0}^{N-1}\sum_{l=0}^1\,
 \int_\FF
\frac{d^2\tau}{\tau_2} \sum_{m_1, m_2, n_2\in \zzz,
n_1\in \zzz+{r\over N}, b\in 2\zzz + l}
\exp\left[ 2\pi i \tau( m_1 n_1 + m_2 n_2 +\frac{b^2}{4} ) \right]
\times \cr
&\;& \;\;\; \exp \left(\frac{-\pi \tau_2}{Y} \left|
n_2 ( \ws \wrh  -\wv ^2) + b\wv  + n_1 \ws  -\wrh m_1 + m_2 \right|^2 \right)\,
e^{2\pi i m_1 s / N} \, h^{(r,s)}_l(\tau)\, , \nonumber \\
\een
where
\be{eu1a}
\Omega=\pmatrix{ \ws  & \wv \cr \wv  & \wrh }\, , \qquad Y = \det
Im\Omega\, ,
\ee
and
\bea{eu2}
h^{(r,s)}_0(\tau) &=& {1\over 2\vt_3(2\tau,2z)} \left( 
F^{(r,s)}(\tau, z) +  F^{(r,s)}(\tau, z+{1\over 2}) \right)\, , \nonumber \\
h^{(r,s)}_1(\tau) &=& {1\over 2\vt_2(2\tau,2z)} \left( 
F^{(r,s)}(\tau, z) -  F^{(r,s)}(\tau, z+{1\over 2}) \right)\, .
\eea
The functions $F^{(r,s)}(\tau, z)$ have been 
defined in eqs.\refb{fifth}.
The coefficients $c^{(r,s)}(4n)$ introduced in \refb{esi4b} are
related to $h^{(r,s)}_l(\tau)$ through the expansion:
\be{eu6a}
h^{(r,s)}_0(\tau) = \sum_{n\in \zzz/N } 
c^{(r,s)}(4n) 
q^n\, , \qquad h^{(r,s)}_1(\tau)
= \sum_{n\in \zzz/N-{1\over 4}  } 
c^{(r,s)}(4n) q^n\, , \qquad q\equiv e^{2\pi i \tau}\, .
\ee
{}From eqs.\refb{fifth}-\refb{esi4b} it follows that
the only non-zero $c^{(r,s)}(4n)$ with negative $n$
are $c^{(r,s)}(-1)$ for all $N$,
and $c^{(r,s)}(-1+{4\over N})$ for
$N=5,7$. They have the values:
\bea{eu6b}
c^{(0,s)}(-1) &=& {2\over N}\, , \nonumber \\
c^{(r,s)}(-1) &=& 0 \qquad \hbox{for $r\ne 0$ mod $N$}\, ,
\nonumber \\
c^{(0,s)}(-1+{4\over N}) &=& 0\, , \nonumber \\
c^{(r,rk)}(-1+{4\over N}) &=& -{48\over N (N+1) (N-1)} \,
e^{2\pi i k/N}\, ,
 \qquad \hbox{for $r\ne 0$ mod $N$}\, .
\een

Eq.\refb{eu1} shows that the zeroes and poles of $\wt\Phi_k$ appear only
when the $\tau$ integral on the right hand side of this equation diverges 
from the region near $\tau=i\infty$. 
Now, if we consider a term proportional to $e^{2\pi i n\tau}$
in the expansion of $h_l^{(r,s)}$, then for large $\tau_2$,
the $\tau_1$ integral gives a
non-vanishing answer only if
\be{eu8}
n + m_1 n_1 + m_2 n_2 +\frac{b^2}{4} = 0\, .
\ee
Thus after performing the $\tau_1$ integral, 
the only $\tau_2$ dependence of the integrand
in the large $\tau_2$ region
comes from the
\be{eu8a}
- \frac{\pi \tau_2}{Y} \left|
n_2 ( \ws \wrh  -\wv ^2) + b\wv  + n_1 \ws  -\wrh m_1 + m_2 \right|^2  
\ee
factor in the exponent.  Since this is negative, the only way the
integral can diverge from the large $\tau_2$ region is if this vanishes:
 \be{eu7}
n_2 ( \ws \wrh  -\wv ^2) + b\wv  + n_1 \ws  -\wrh m_1 + m_2 =0\, , 
\ee
for some $m_1$, $m_2$, $n_1$, $n_2$, $b$ appearing in the sum in 
\refb{eu1}.  

Now  we have the identity
\be{eu7a}
 m_1 n_1 + m_2 n_2 +\frac{b^2}{4} ={1\over 2} (p_L^2 - p_R^2)\, ,
 \ee
 where
 \bea{eu7aa}
 p_R^2 &=& {1\over 2 Y} \left|
n_2 ( \ws \wrh  -\wv ^2) + b\wv  + n_1 \ws  -\wrh m_1 + m_2 \right|^2, \nonumber \\
p_L^2 &=& {1\over 2Y} \left\{ m_2 + n_2(\ws_2\wrh_2+\ws_1\wrh_1-\wv_1^2-\wv_2^2) - m_1 \wrh_1 + n_1 \ws_1 
+ b \wv_1\right\}^2 \nonumber \\
&& + {1\over 2 Y} \left\{ n_2 \left(\ws_1\wrh_1 - \ws_1 \wrh_2 
+ 2 \wv_1\wv_2 - {2\wv_2^2\wrh_1\over
\wrh_2}\right) + m_1\wrh_2
+ n_1 \left(\ws_2 - 
{2 \wv_2^2\over \wrh_2}\right) - b\wv_2\right\}^2\nonumber \\
&& + 2 \left\{ {b\over 2} + n_1 {\wv_2\over \wrh_2} - n_2\wv_1
+ n_2 {\wv_2 \wrh_1\over \wrh_2}\right\}^2\, .
\een
 Since $p_L^2$ is positive semi-definite,  and
since $p_R^2$ vanishes when \refb{eu7} holds,
\refb{eu7a}
 shows that
we must have
 \be{eu7b}
 m_1 n_1 + m_2 n_2 +\frac{b^2}{4}  \ge 0\, .
 \ee
 Furthermore the equality sign holds only when $p_L^2$ also
 vanishes. This requires $m_1=m_2=n_1
 =n_2=b=0$. The corresponding divergence is present for all
 $\ws $, $\wrh $, $\wv $ and is removed by a subtraction 
 term\cite{0602254}. 
 Thus the
 divergences which depend on $\ws $, $\wrh $, $\wv $ come from those
 values of $m_i$, $n_i$, $b$ which satisfy \refb{eu7}
 and for which
 \be{eu7ba}
 m_1 n_1 + m_2 n_2 +\frac{b^2}{4}  > 0\, .
 \ee
 This, together with
 eq.\refb{eu8}, now show that we must have
 \be{eu7c}
 n < 0\, .
 \ee
 In other words the only terms in the expansion of $h_l^{(r,s)}$ 
 responsible for a divergent contribution to the integral 
 \refb{eu1} are the ones involving negative powers of $q$. {}From
eqs.\refb{eu6a}, \refb{eu6b} it now follows that for all $N$ there
is a
divergent
contribution to \refb{eu1}  from the $n=-1/4$
 term, \i.e. the term
\be{ev1}
c^{(r,s)}(-1) q^{-1/4}\, ,
\ee
in $h^{(r,s)}_1$. On the other hand for $N=5,7$ there are additional
divergences from $n=-{1\over 4}+{1\over N}$, \i.e.
the term 
\be{ev1x}
c^{(r,s)}(-1+{4\over N}) q^{-1/4+1/N}\, ,
\ee
in $h^{(r,s)}_1$.
Since the subscript of $h$ is odd in both cases, we must have 
$b$  odd.

First consider the contribution from the $c^{(r,s)}(-1)$ term.
\refb{eu8} now gives
\be{ev2}
m_1 n_1 + m_2 n_2 +\frac{b^2}{4} = {1\over 4}\, .
\ee
After estimating the $\tau_2$ integral in \refb{eu1}
for $n_2 ( \ws \wrh  -\wv ^2) + 
b\wv  + n_1 \ws  -\wrh m_1 + m_2\simeq 0$, one easily finds that
the 
divergent contribution is given by
\bea{ev2a}
-2\sum_{s=0}^{N-1}
e^{2\pi i m_1 s / N}\, c^{(r,s)}(-1)\, 
\ln \left|
n_2 ( \ws \wrh  -\wv ^2) + b\wv  + n_1 \ws  -\wrh m_1 + m_2 \right|^2\, ,
\nonumber \\
\quad \hbox{$r=n_1N$ mod $N$, $b=1$ mod 2}\, , \quad
\een
where we have included a factor of 2 due to the fact that  
the lattice vectors
$(\vec m, \vec n, b)$ and $(-\vec m, -\vec n, -b)$
give identical divergent
contribution.
Thus 
near this region $\wt\Phi_k$ behaves as
\be{ev3}
\wt\Phi_k \sim \left( n_2 ( \ws \wrh  -\wv ^2) + b\wv  + 
n_1 \ws  -\wrh m_1 + m_2
\right)^{\sum_{s=0}^{N-1}
e^{2\pi i m_1 s / N}\, c^{(r,s)}(-1)}\, .
\ee
Using \refb{eu6b} we can compute the sum in the
exponent. In particular $r$ must vanish for $c^{(r,s)}(-1)$
to be non-zero, and hence
$n_1$ must be an integer. Furthermore, 
substituting $c^{(0,s)}(-1)=2/N$
in \refb{ev3} we see that the sum in the exponent 
vanishes unless $m_1$ is an integer multiple
of $N$. The final result after performing the sum is
\bea{ev4}
\wt\Phi_k &\sim& \left( n_2 ( \ws \wrh  -\wv ^2) + b\wv  
+ n_1 \ws  -\wrh m_1 + m_2
\right)^2 \nonumber \\
&& \quad  \hbox{for $m_1\in N\ZZZ$, $n_1\in\ZZZ$,
$b\in 2\ZZZ+1$, $m_2, n_2\in \ZZZ$}\, ,
\nonumber \\
&\sim& 1 \quad \hbox{otherwise}\, .
\een

For $N=5,7$ we also have divergent contribution to \refb{eu1}
from the
$c^{(r,s)}(-1+{4\over N})$ term. 
In this case \refb{eu8}   gives
\be{ev2x}
m_1 n_1 + m_2 n_2 +\frac{b^2}{4} = {1\over 4}-{1\over N}\, .
\ee
The divergent contribution takes the form
\bea{ev2aa}
-2\sum_{s=0}^{N-1}
e^{2\pi i m_1 s / N}\, c^{(r,s)}(-1+{4\over N})\, 
\ln \left|
n_2 ( \ws \wrh  -\wv ^2) + b\wv  + n_1 \ws  -\wrh m_1 + m_2 \right|^2\, ,
\nonumber \\
\quad \hbox{$r=n_1N$ mod $N$, $b=1$ mod 2}\, . \quad
\een
Thus 
$\wt\Phi_k$ behaves as
\be{ev3aa}
\wt\Phi_k \sim \left( n_2 ( \ws \wrh  -\wv ^2) + b\wv  + n_1 \ws  
-\wrh m_1 + m_2
\right)^{\sum_{s=0}^{N-1}
e^{2\pi i m_1 s / N}\, c^{(r,s)}(-1+{4\over N})}\, .
\ee
We can compute the sum in the
exponent by rewriting it as
\be{eex1}
\sum_{s=0}^{N-1}
e^{2\pi i m_1 s / N}\, c^{(r,s)}(-1+{4\over N})
=\sum_{k=0}^{N-1} e^{2\pi i m_1 rk / N}
\, c^{(r,rk)}(-1+{4\over N})  \quad \hbox{for $r\ne 0$ mod $N$}\, .
\ee
Using \refb{eu6b} we now get
\be{eex2}
\sum_{s=0}^{N-1}
e^{2\pi i m_1 s / N}\, c^{(r,s)}(-1+{4\over N}) 
= \cases{-48/(N^2-1) \quad \hbox{for $m_1 r= -1$ 
mod $N$} \cr 0 \quad \hbox{otherwise} }\, .  
\ee
This gives
\bea{ev4aa}
\wt\Phi_k &\sim& \left( n_2 ( \ws \wrh  -\wv ^2) + b\wv  
+ n_1 \ws  -\wrh m_1 + m_2
\right)^{-48/(N^2-1)} \nonumber \\
&& \quad  \hbox{for $m_1 n_1 N=-1$ mod $N$, $n_1N\ne 0$ mod $N$,
$b\in 2\ZZZ+1$, $m_1,m_2,n_2\in \ZZZ$}\, ,
\nonumber \\
&\sim& 1 \quad \hbox{otherwise}\, .
\een

To summarize, for $N=1,2,3,5,7$, $\wt\Phi_k$ has a second order zero
at \refb{eu7} for $m_1=0$ mod $N$, $n_1=0$ mod 1 and odd $b$,
satisfying \refb{ev2}.
On the other hand for $N=5$ and $N=7$, $\wt\Phi_k$ also has poles
of order $48/(N^2-1)$ at \refb{eu7} for $m_1n_1 N=-1$ mod $N$, 
$n_1N\ne 0$ mod $N$ and odd $b$, satisfying \refb{ev2x}.
(Note that both for $N=5$ and $N=7$,
$48/(N^2-1)$ is an integer, and hence the singularities are poles and
not branch points.)

\sectiono{Riemann Normal Coordinates and Duality Invariant Statistical
Entropy Function} \label{s3}

In section \ref{sasymp} we considered 
$\htau=\vtau -\vtau_B$ for some fixed base point $\vtau_B$
as the fundamental field
in defining $W_B(\vtau_B,\vec J)$ and $\Gamma_B(\vtau_B, \vc)$. In this
appendix we shall try to generalize this by treating 
\be{ef1}
\vxi = \vec g(\htau)
\ee
as a fundamental field where $\vec g(\htau)$ is an arbitrary function
of $\htau$ with a Taylor series expansion starting with the linear terms
(\i.e. $\vec g(\vec 0)=\vec 0$). In this case the generating function
of $\vxi$ correlation functions will be given by
\be{ef2}
e^{\wt W_B(\vtau_B,\vec J)} = \int{d^2\eta\over (\tau_{B2}+\eta_2)^2} 
\, e^{-F(\vtau_B+\htau) + \vec J\cdot
\vec g(\htau)}\, .
\ee
Thus $\wt W_B(\vtau_B,\vec 0)=S_{stat}$.
The corresponding effective action is
\be{ef3}
\wt \Gamma_B(\vtau_B, \vpsi) =  \vec J\cdot \vpsi - 
\wt W_B(\vtau_B,\vec J) \, ,
\qquad \psi_i = {\p \wt W_B(\vtau_B,\vec J)\over \p J_i}\, .
\ee
{}From this definition it follows that
\be{ef3a}
J_i = {\p \wt \Gamma_B(\vtau_B, \vpsi)\over \p \psi_i}\, .
\ee

Now suppose $\vtau^{(0)}$ is a specific value of $\vtau_B$ for which
\be{ef4}
{\p \wt
\Gamma_B(\vtau^{(0)}, \vpsi)\over \p\psi_i}\bigg|_{\vpsi =0}
= 0 \quad  \i.e. \quad {\p \wt W_B(\vtau^{(0)},\vec J)
\over \p J_i}\bigg|_{\vj=0}=0\, .
\ee
In this case we have $\vec J=0$ for $\vpsi=0$, and hence
\be{ef5}
\wt\Gamma_B(\vtau^{(0)}, \vec 0) = - \wt W_B(\vtau^{(0)},\vec 0)
= - S_{stat}\, .
\ee
We shall now show that $\wt\Gamma_B(\vec \tau_B, \vec 0)$, regarded
as a function of $\vec\tau_B$, has an 
extremum at $\vtau_B=\vtau^{(0)}$. From \refb{ef3}, \refb{ef3a}
we see
that
\be{ef6}
\wt\Gamma_B(\vtau^{(0)}+\vec\epsilon, \vec 0)
= - 
\wt W_B\left.\left(\vtau^{(0)}+\vec\epsilon,
\vec J=\p \wt\Gamma_B(\vtau^{(0)}+
\vec\epsilon, \vec\psi) / \p\vec \psi\right)
\right|_{\vpsi=\vec 0} \, .
\ee
Now
\be{ef7}
e^{\wt W_B(\vtau_B +\vec\epsilon,
\vec J)} = \int{d^2\eta\over (\tau_{B2}+\epsilon_2+
\eta_2)^2} 
\, e^{-F(\vtau_B+\vec\epsilon+\htau) + \vec J\cdot
\vec g(\htau)}
= \int{d^2\eta\over (\tau_{B2}+\eta_2)^2} 
\, e^{-F(\vtau_B+\htau) + \vec J\cdot
\vec g(\htau-\vec\epsilon)}\, ,
\ee
where in the second step we have made a change of variables
$\vec\eta\to \vec\eta-\vec\epsilon$. Since $g(\htau-\vec\epsilon)
= g(\htau) + O(\vec\epsilon)$, this shows that
\be{ef8}
 \wt W_B(\vtau_B +\vec\epsilon,
\vec J) = \wt W_B(\vtau_B ,
\vec J) + O(\eps_i J_k)\, .
\ee
Using this information in \refb{ef6} we get
\be{ef9}
\wt\Gamma_B(\vtau^{(0)}+\vec\epsilon, \vec 0)
= - 
\wt W_B(\vtau^{(0)},
\vec J=\p \wt\Gamma_B(\vtau^{(0)}+
\vec\epsilon, \vec\psi) / \p\vec \psi)|_{\vpsi=\vec 0}
+ O\left(\epsilon_i {\p \wt\Gamma_B(\vtau^{(0)}+
\vec\epsilon, \vec\psi)\over \p \psi_i}\bigg|_{\vpsi=0}\right)
\, .
\ee
Eq.\refb{ef4} shows that the second term on the right hand side
of this equation is of order
$\eps^2$, and $\vec J$ appearing in the argument of the first term
is of order $\eps$. Expanding the first term in a Taylor series 
expansion in $\vj$, and noting that the linear term vanishes
due to \refb{ef4}, we get
\be{ef10}
\wt\Gamma_B(\vtau^{(0)}+\vec\epsilon, \vec 0)
= - 
\wt W_B(\vtau^{(0)},
\vec J=\vec 0)  
+O(\epsilon^2) = \wt\Gamma_B(\vtau^{(0)}, \vec 0) 
+O(\epsilon^2)\, .
\ee
Thus
\be{ef11}
{\p \wt\Gamma_B(\vtau, \vec 0)\over \p \tau_i}=0 \quad \hbox{at}
\quad \vec \tau = \vtau^{(0)}\, .
\ee
Using \refb{ef5} and \refb{ef11} we see that the statistical entropy is
given by the value of
$-\wt\Gamma_B(\vtau, \vec 0)$ at its extremum $\vtau = \vtau^{(0)}$.
Thus we can identify $-\wt\Gamma_B(\vtau, \vec 0)$ as the 
statistical entropy
function. This is computed as the sum of
1PI vacuum amplitudes in the theory
with $\xi_i$ regarded as the fundamental fields.

We shall now show that for a suitable choice of the coordinates
$\vxi$, the statistical
entropy function $-\wt\Gamma_B(\vec\tau, \vec 0)$
defined this way can be made manifestly duality invariant. This is done
by choosing $\vec\xi$ as Riemann normal coordinates. For a given
base point $\vtau_B$ we define the coordinate $\vxi$ for a given
point $\vtau$ in the upper half plane as follows. We introduce the duality
invariant metric on the upper half plane
\be{eg1}
ds^2 = (\tau_2)^{-2} (d\tau_1^2 + d\tau_2^2)\, ,
\ee
and draw a geodesic connecting $\vtau_B$ and $\vtau$. The coordinate
$\vxi$ corresponding to the point $\vtau$ is then defined by identifying
$|\vxi|$ as the distance between $\vec\tau_B$ and $\vtau$ along the
geodesic and the direction of $\vxi$ is taken to be the direction of
the tangent vector to the geodesic at $\vtau_B$.\footnote{Often one
uses the convention that the distance along the geodesic is
$\sqrt{g_{ij}(\vec\tau_B) \xi^i\xi^j}$. This definition differs 
from the one used here by a multiplicative factor of $\tau_{2B}$.
Since this transforms covariantly under a duality transformation,
both choices of $\vec\xi$ would give manifestly covariant Feynman
rules.} 
Since the metric 
\refb{eg1} is
invariant under a duality transformation, it is clear that if a duality
transformation maps $\vtau_B$ to $\vtau_B'$ and $\vtau$ to $\vtau'$,
then the Riemann normal coordinate $\vxi'$ of $\vtau'$ with respect
to $\vtau'_B$ will have the property that $|\vxi'|=|\vxi|$. Thus $\vxi$
and $\vxi'$ are related by a rotation. In order to determine the angle
of rotation, we note that under a duality transformation
\refb{egamman},
\be{eg2}
d\tau' = (c\tau + d)^{-2} d\tau\, .
\ee
Thus
\be{eg3}
{d\tau'\over |d\tau'|} = {|c\tau + d|^2\over (c\tau+d)^2}\, 
{d\tau\over |d\tau|}\, .
\ee
This shows that a geodesic passing through $\tau_B$ gets rotated by
a phase $|c\tau_B + d|^2/(c\tau_B+d)^2$ under 
a duality transformation. Hence
\be{eg4}
\xi' = {|c\tau_B + d|^2\over (c\tau_B+d)^2}\, \xi = 
{c\bar\tau_B + d \over c\tau_B+d }\, \xi \, ,
\ee
where
\be{eg5}
\xi = \xi_1 + i \xi_2, \qquad \xi' = \xi'_1 + i \xi'_2\, .
\ee

Since for given $\tau_B$ the duality transformation acts linearly
on $\vxi$, the corresponding generating function $\wt W_B(\vtau_B,
\vj)$ and the effective action $\wt\Gamma_B(\vtau_B, \vpsi)$ will
be manifestly duality invariant under simultaneous transformation
of $\vtau_B$, $\vj$ or $\vpsi$ and of course the charges $Q_e$
and $Q_m$. In particular the 1PI vacuum amplitude 
$\wt\Gamma_B(\vtau, \vec 0)$ will be duality invariant under
the transformation \refb{egamman}.

We shall now give an algorithm for explicitly generating duality
covariant vertices in this 0-dimensional field theory. For this we
need to expand the duality invariant function $F(\vtau)$ in a Taylor
series expansion in $\vxi$. This is given by:
\be{eh1}
F(\vtau) = \sum_{n=0}^\infty {1\over n!} (\tau_{B2})^n
\xi_{i_1}\ldots
\xi_{i_n}\, D_{i_1}
\cdots D_{i_n} F(\vtau)\bigg|_{\vtau=\vtau_B}\, ,
\ee
where $D_i$ denotes  covariant derivative with respect to $\tau_i$,
computed using  the
affine connection $\Gamma^i_{jk}$ constructed from the 
metric \refb{eg1}. We arrive at \refb{eh1} by using the result that
in the $\vec\xi$ coordinate system symmetrized covariant derivatives
are equal to ordinary derivatives. Using this we can replace ordinary
derivatives in the Taylor series expansion to covariant derivatives with
respect to $\xi_i$. In the second step we use the tensorial transformation
properties of covariant derivatives to convert covariant derivative with
respect to $\xi_i$ to covariant derivative with respect to $\tau_i$. The
$(\tau_{B2})^n$ factor in \refb{eh1} arises due to the fact that near
$\vtau=\vtau_B$,
\be{eh1.9}
d\tau_i = \tau_{B2} d\xi_i\, .
\ee

In the $(\tau,\bar\tau)$ coordinate system
the non-zero components of the connection
are
\be{eh2}
\Gamma^\tau_{\tau\tau} = {i\over \tau_2}, \qquad
\Gamma^{\bar\tau}_{\bar\tau\bar\tau} = -{i\over \tau_2}\, .
\ee
The  curvature tensor computed from this connection has the form
\be{eh3}
R^i_{~jkl} = - (\delta^i_k g_{jl} - \delta^i_l g_{jk})\, ,
\ee
which shows that the metric \refb{eg1} describes a constant negative
curvature metric. From \refb{eh2} it follows that
\bea{eh4}
D_\tau (D_\tau^m D_{\bar\tau}^n F(\vec\tau))
&=& (\p_\tau - im/\tau_2) (D_\tau^m D_{\bar\tau}^n F(\vec\tau)),
\nonumber \\
D_{\bar\tau} (D_\tau^m D_{\bar\tau}^n F(\vec\tau))
&=& (\p_{\bar\tau} + in/\tau_2)
(D_\tau^m D_{\bar\tau}^n F(\vec\tau))\, ,
\een
for any arbitrary ordering of $D_\tau$ and $D_{\bar\tau}$ 
in $D_\tau^m D_{\bar\tau}^n F(\vec\tau)$. 
\refb{eh4} provides us with explicit expressions for the
covariant derivatives of $F$ appearing in \refb{eh1}. Also
using
\refb{eh4} one can prove iteratively that under a duality
transformation
\be{eh5}
(\tau_2)^{m+n} \, D_\tau^m D_{\bar\tau}^n F(\vec\tau)
\to \left({c\tau+d\over c\bar\tau+d}\right)^{m-n}\,
 (\tau_2)^{m+n} \, D_\tau^m D_{\bar\tau}^n F(\vec\tau)\, ,
 \ee
 under a duality transformation. This establishes manifest
 duality covariance of the vertices constructed from
 \refb{eh1}.
 
 We also need to worry about the contribution from the integration
 measure. The original measure $d^2\tau/(\tau_2)^2$ was duality
 invariant. Since duality transformation acts on $\vec\xi$ as a rotation,
 $d^2\xi$ is also a duality invariant measure. Thus we must have
 \be{eh5x}
 {d^2\tau\over (\tau_2)^2} = \JJ(\tau_B, \vxi) \, d^2\xi\, ,
 \ee
 for some duality invariant function $\JJ(\tau_B,\vxi)$. It has been
 shown in appendix \ref{sa} that
 \be{eh6}
 \JJ(\tau_B,\vxi) = {1\over |\vec\xi|} \sinh{|\vec\xi|}\, .
 \ee
 We can now regard $-\ln \JJ(\tau_B,\vxi)$ as an additional contribution
 to the action and expand this in a power series expansion in $\vxi$
 to generate additional vertices. Using the expression for $F(\vec\tau)$
 given in \refb{ek2} we now see that the full `action' is given by
 \bea{eh7}
 F(\vtau) -\ln \JJ(\tau_B,\vxi)
 &=&
  -\Bigg[ {\pi\over 2 \tau_2} \, |Q_e +\tau Q_m|^2
-\ln f^{(k)}(\tau) -\ln f^{(k)}(-\bar\tau) - (k+2) \ln (2\tau_2)
\nonumber \\
&& +\ln\bigg\{ K_0 {\pi\over \tau_2} |Q_e +\tau Q_m|^2 
\bigg\} + \ln \JJ(\tau_B,\vxi) \nonumber \\
&& + \ln\left( 1 + { 
2(k+3) \tau_2 \over \pi  |Q_e +\tau Q_m|^2 }\right)\Bigg] \, .
 \nonumber \\
\een
In this expression the first term inside the square
bracket is quadratic in the charges, the last term contains terms
of order $Q^{-2n}$ for $n\ge 1$, and
the other terms are of order $Q^0$. Thus in order to carry out a
systematic expansion in powers of inverse charges we need to regard
the first term as the tree level contribution, the last term as two and
higher loop contributions and the other terms as
the 1-loop contribution. We can now evaluate the effective action 
$\wt\Gamma_B(\vec\tau_B)$ in a systematic loop expansion.
The leading term in the effective action is then just the first term in 
\refb{eh7} evaluated at $\vtau=\vtau_B$:
\be{eh8}
\wt\Gamma_0(\tau_B) = -{\pi\over 2 \tau_{2B}} \, |Q_e +\tau_B Q_m|^2
\, .
\ee
At the next order we need to include the tree level contribution from the
rest of the terms in the action (except the last term which is higher
order) and one loop
contribution from the first term. 
The former corresponds to these
terms being evaluated at $\vtau=\vtau_B$, \i.e. $\vec\xi=0$.
Since $\JJ(\vtau_B, \vxi=0)=1$, we get this contribution to be
\be{ei0}
\ln f^{(k)}(\tau_B) +\ln f^{(k)}(-\bar\tau_B) + (k+2) \ln (2\tau_{2B})
- \ln\bigg\{K_0\, {\pi\over \tau_{2B}} |Q_e +\tau_B Q_m|^2\big)
\bigg\}\, .
\ee
For the one loop contribution from the first term in the action
we need to expand this term
to quadratic order in $\vxi$ using eqs.\refb{eh1}, \refb{eh4}.
The
order $\vxi$ and $\xi^2$ terms are given by
\be{ei1}
-{i\pi\over 4\tau_{2B}} \left\{ \bar\xi (Q_e+Q_m\tau_B)^2
+ \xi (Q_e+Q_m\bar\tau_B)^2 \right\}
-{\pi\over 4 \tau_{2B}} \, |Q_e +\tau_B Q_m|^2 \,
\bar\xi \xi\, .
\ee
The linear term in $\vxi$ do not give any contribution to the 1PI
amplitudes. The contribution from the quadratic term gives
\be{ei2}
\ln \left({1\over 4 \tau_{2B}} \, |Q_e +\tau_B Q_m|^2\right)\, .
\ee
Thus the complete one loop contribution to the effective action
is given by
\be{ei4}
\wt \Gamma_1(\tau_B) = \ln f^{(k)}(\tau_B) +\ln f^{(k)}(-\bar\tau_B) 
+ (k+2) \ln (2\tau_{2B}) -\ln (4 \pi K_0) \, .
\ee
Up to an additive constant this agrees with the black hole entropy 
function for CHL models
calculated in \cite{0508042}.

 \sectiono{The Integration Measure $\JJ(\vec\tau_B, \vec\xi)$} 
 \label{sa}
 
 In this appendix we shall compute the integration measure 
 $\JJ(\vec\tau_B,\vec\xi)$ which arises from a change of variables
 from $\tau_1,\tau_2$ to the normal coordinates:
 \be{eh5xapp}
 {d^2\tau\over (\tau_2)^2} = \JJ(\vec\tau_B, \vxi) \, d^2\xi\, .
 \ee
 We first note that the duality invariant metric
 \be{esa1}
 {1\over \tau_2^2} ( d\tau_1^2 + d\tau_2^2)
 \ee
 describes a metric of constant negative curvature $-1$. Since this is
 a homogeneous space, $\JJ(\vtau_B,\vxi)$ cannot depend on
 $\vec\tau_B$. Furthermore
 with
 an appropriate change of variables we can bring the metric and measure 
 to the
 standard form:
 \be{esa2}
 ds^2 = d\theta^2 + \sinh^2\theta d\phi^2\, , \qquad
 {d^2\tau\over \tau_2^2} = \sinh\theta \, d\theta\, d\phi\, .
 \ee
 Since $\JJ$ is independent of the base point $\vtau_B$, 
 we can calculate it by taking
 the base point to be at $\theta=0$. The geodesics passing through
 this point are constant $\phi$ lines, and the length measured along
 such a geodesic is given by $\theta$. Thus we have
 \be{esa3}
 \vxi = (\theta\cos\phi, \theta\sin\phi)\, .
 \ee
 This gives 
 \be{esa4}
 d^2\xi = \theta d\theta d\phi\, .
 \ee
 Comparing this with \refb{esa2} we get
 \be{esa4a}
 {d^2\tau\over \tau_2^2} = {\sinh\theta\over \theta} d^2\xi
 = {1\over |\vxi|} \sinh|\vxi| \, d^2\xi\, .
 \ee
 Thus
 \be{esa5}
 \JJ(\vtau_B,\vxi) = {1\over |\vxi|} \sinh|\vxi|\, .
 \ee

\end{document}